# Bayesian mixture models for phylogenetic source attribution from consensus sequences and time since infection estimates.

Alexandra Blenkinsop[1*], Lysandros Sofocleous[1], Francesco Di Lauro[2], Evangelia Georgia Kostaki[3], Ard van Sighem[4], Daniela Bezemer[4], Thijs van de Laar[5], Peter Reiss[6,7], Godelieve de Bree[6,8], Nikos Pantazis[2], and Oliver Ratmann[1]
on behalf of the HIV Transmission Elimination Amsterdam Initiative

[1]Department of Mathematics, Imperial College London, UK
[2]Big Data Institute, Nuffield Department of Medicine, University of Oxford, UK
[3]Department of Hygiene, Epidemiology and Medical Statistics, Medical School, National and Kapodistrian University of Athens, Greece
[4]Stichting HIV Monitoring, Netherlands
[5]Department of Donor Medicine Research, Sanquin, Netherlands
[6]Amsterdam Institute for Global Health and Development, Netherlands
[7]Department of Global Health, Amsterdam University Medical Centers, Netherlands
[8]Division of Infectious Diseases, Department of Internal Medicine, Amsterdam Infection and Immunity Institute, Netherlands
[*]a.blenkinsop@imperial.ac.uk

**Abstract**

In stopping the spread of infectious diseases, pathogen genomic data can be used to reconstruct transmission events and characterize population-level sources of infection. Most approaches for identifying transmission pairs do not account for the time passing since divergence of pathogen variants in individuals, which is problematic in viruses with high within-host evolutionary rates. This prompted us to consider possible transmission pairs in terms of phylogenetic data and additional estimates of time since infection derived from clinical biomarkers. We develop Bayesian mixture models with an evolutionary clock as signal component and additional mixed effects or covariate random functions describing the mixing weights to classify potential pairs into likely and unlikely transmission pairs. We demonstrate that although sources cannot be identified at the individual level with certainty, even with the additional data on time elapsed, inferences into the population-level sources of transmission are possible, and more accurate than using only phylogenetic data without time since infection estimates. We apply the approach to estimate age-specific sources of HIV infection in Amsterdam MSM transmission networks between 2010-2021. This study demonstrates that infection time estimates provide informative data to characterize transmission sources, and shows how phylogenetic source attribution can then be done with multi-dimensional mixture models.

## 1 Introduction

Genomic surveillance of human pathogens is increasingly used to help combat the spread of infectious diseases such as COVID-19, antimicrobial resistant bacteria, Ebola virus, or HIV[1–5]. This involves the sequencing of the genetic code of pathogen samples obtained from diagnosed individuals[6] or the environment such as wastewater[7,8], and then phylogenetic analyses are used to estimate ancestral relationships between samples based on an assumed model which describes the rate of mutation as pathogens evolve[9]. Inferred evolutionary relationships are represented by phylogenetic trees, in which branch lengths represent the degree of genetic variation between samples which appear as tips of the tree, and these phylogenetic trees are indicative of epidemic transmission networks among human hosts[10]. These methods can be used for example to detect new circulating pathogens or pathogen variants[11], determine growth rates[12], quantify modes of disease spread[13], or characterize population-level drug resistance[14,15]. Particular interest centers on reconstructing pathogen transmission, with the primary aim to identify population-level factors that underpin disease spread[16,17]. It is usually not possible to determine with certainty from the genetic data alone that one individual is the source of infection in another person, particularly in the case of fast-evolving pathogens such as HIV. For instance, it is common to observe genetically near-identical HIV sequences between women, even though HIV transmission between women is highly unlikely, and molecular patterns of near-identical virus suggest instead that one or more men of the same transmission chain remained unobserved[18]. For this reason, population-level inferences into the drivers of pathogen transmission focus on analyses that seek to harness the information contained in phylogenetic trees spanning all available samples[19–22], particular parts of phylogenetic trees[23,24], or a larger number of phylogenetically reconstructed transmission pairs[17,25,26]. The latter approaches have proven particularly useful when additional data provides insights into the direction of transmission[10,27], as then flexible and computationally efficient regression methods using attributes of the likely source and recipient can be used to quantify transmission flows[28,29].

Large sets of phylogenetic transmission pairs are typically identified using genetic distances between pathogen sequences or patristic distances along lineages in phylogenetic trees, sometimes coupled with additional criteria including the statistical support that the two individuals are part of the same sub-tree of the true, unknown phylogeny, or the depth of the lineage separating the two individuals, often expressed in units of calendar time[30]. In practice, these linkage criteria are often loosely justified, but especially those based on evolutionary distances can be more firmly grounded in statistical models on the expected number of genetic mutations under a generative evolutionary clock model[31,32]. Clock models describe genetic distance between two sequences in terms of the amount of time elapsed since the lineages leading to the two observed sequences diverged and are used widely to characterize pathogen evolution, including for example the evolution of novel severe acute respiratory syndrome coronavirus 2 (SARS-CoV-2) variants in immuno-compromised patients[33]. The challenge is to define the time elapsed, since the divergence time of the lineages leading to the observed sequences is unknown. We address this, using HIV as an example, by leveraging additional data that can be used to estimate time since infection, and approximate the time elapsed as the interval between the infection event and the sampling time of two observed sequences, which assumes that the two lineages diverged at the infection event (Figure 1). We then express the likelihood that observed genetic distances evolved within a specified time elapsed between two individuals under a standard evolutionary clock model. This approach can account for substantial natural heterogeneity in time elapsed when attempting to interpret genetic distances arising from variation in when HIV is transmitted, how late individuals are diagnosed, and when diagnosed individuals' samples are sequenced. We then embed this transmission pair likelihood into a two-component Bayesian mixture model to estimate posterior probabilities that pairs of individuals are actual transmission pairs, relative to a background noise distribution on genetic distances and time elapsed. We will see that the posterior transmission pair probabilities are themselves uncertain and so are not of immediate interest, but by aggregating the posterior transmission pair probabilities we can identify and quantify the drivers of transmission at population-level.

Bayesian mixture models (BMMs) are widely used to classify points into latent components of similar data[34], characterized by a distinct probability distributions. As such, BMMs are a natural starting point for interpreting two-dimensional point patterns of genetic distances and time elapsed. We limit ourselves to allocating points probabilistically into two components —a signal and a background component— rather than exploring high or infinite-dimensional components[35], as our primary interest centers on the population-level drivers of infection and not the point classifications. More fundamentally, unrelated pairs of individuals can have genetic distances and times elapsed that are compatible with the HIV evolutionary rate. This is a challenge because the number of all possible pairwise combinations of individuals far exceeds the number of transmission pairs. We adopt previous work which attaches generalized linear predictors to the BMM mixing weights to improve classification accuracy[36] and other big data reduction techniques, circumventing highly imbalanced classification problems.

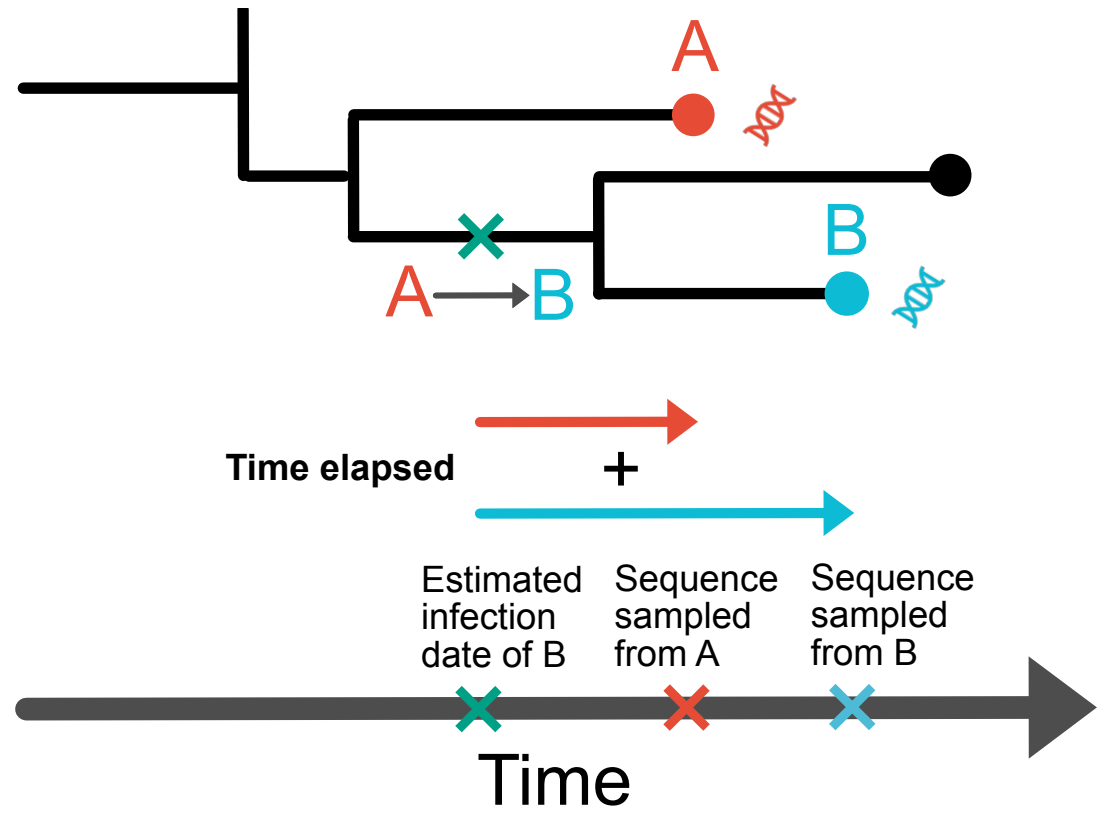

Figure 1: **Schematic of a phylogenetic tree illustrating how to approximate time elapsed.** We consider that A transmitted to B, with the sequence sampling dates of A and B known. We estimate the infection date of individual B, and estimate the time elapsed as the cumulative time between the infection date of B and each of the sequence sampling dates of A and B.

Sections 2.1 to 2.3 lay out the notation that we use to describe unknown transmission events and transmission flows, and the data available to estimate these. Section 2.4 introduces the Bayesian hierarchical evolutionary clock model that we use to represent the relationship between viral HIV sequences and the time elapsed between them. In Sections 2.5 to 2.7, we integrate the evolutionary clock model as signal component of increasingly complex Bayesian mixture models, and describe how we estimate population-level transmission flows and population-level sources of transmission. Section 3 assesses the performance of the mixture model on simulated data, characterizes the accuracy of phylogenetic source attribution with Bayesian mixture models, and presents techniques to improve classification accuracy when the number of unrelated pairs far exceeds the number of actual transmission pairs. In Section 3.5, we apply the method to characterize the age-specific drivers of HIV transmission between 2010 and 2021 among men having sex with men (MSM) in Amsterdam in the Netherlands, and who were part of phylogenetically reconstructed transmission chains. Finally, Section 4 summarises our findings and discusses the application of our method to other pathogens.

# 2 Methods

## 2.1 Target quantities.

Consider a population of $n$ individuals infected with a pathogen. Of these, $m$ new cases were acquired in a specific time period $\mathcal{T}$, where $m \leq n$. The estimated infection date of an individual $i$, is denoted by $T_i$. We are interested in characterizing the transmission events to the $m$ new cases in $\mathcal{T}$, and denote these with $Z_{ij} = 1$ if $i$ infected $j$ for $i = 1, \dots, n$ and $j = 1, \dots, m$ , and $Z_{ij} = 0$ otherwise . We refer to $i$ as source or transmitting partner of $j$ when $Z_{ij} = 1$, and to $i$ and $j$ as linked when either $Z_{ij} = 1$ or $Z_{ji} = 1$, and to $i$ and $j$ as unlinked if both $Z_{ij} = 0$ and $Z_{ji} = 0$. The unknown $n \times m$ matrix $\boldsymbol{Z}$ is commonly referred to as the adjacency matrix. We primarily wish to characterize population-level transmission flows between groups of individuals specified by distinct demographic, behavioral, or clinical covariates. We denote a partition of the study population with $\mathcal{A}$, and population groups in this partition by $a, b \in \mathcal{A}$, and aggregate over individuals in any of these population groups, which we denote by $i, j \in a$. We quantify the population-level transmission counts in time period $\mathcal{T}$ between population strata with $Z_{ab} = \sum_{i \in a, j \in b} Z_{ij}$ for all $a, b \in \mathcal{A}$. We are primarily interested in the relative contributions of population groups to the transmission counts, and our target quantities are the population-level transmission flows from group $a$ to group $b$, the transmission sources from group $a$ among infections in group $b$, and the transmission sources from group $a$ in the entire population, respectively defined by

$$\pi_{ab} = Z_{ab} \Big/ \Big( \sum_{c,d \in \mathcal{A}} Z_{cd} \Big) \tag{1a}$$

$$\delta_{ab} = Z_{ab} \Big/ \Big( \sum_{c \in \mathcal{A}} Z_{cb} \Big) \tag{1b}$$

$$\delta_{a} = \Big( \sum_{b \in \mathcal{A}} Z_{ab} \Big) \Big/ \Big( \sum_{c,d \in \mathcal{A}} Z_{cd} \Big). \tag{1c}$$

## 2.2 Observations.

A subset of infected individuals are diagnosed, and we denote the number of diagnosed individuals infected with the pathogen by $n^D$ and the number of diagnosed new cases in time period $\mathcal{T}$ by $m^D$. In the Netherlands, all individuals diagnosed with HIV enter the open longitudinal ATHENA cohort except a small proportion of patients who opt out[37]. Using available clinical and demographic biomarker data, the time of infection of diagnosed individuals can be estimated[38,39], which we denote by $\hat{T}_i$ for $i = 1, \dots, n^D$. Then, we consider as potential sources of transmission to a new case all diagnosed individuals with an infection date preceding the case. We denote a “potential source” by

$$Y^D_{ij} = \begin{cases} 1 \text{ if } \hat{T}_i < \hat{T}_j \\ 0 \text{ if } \hat{T}_i \geq \hat{T}_j, \end{cases} \tag{2}$$

where $i = 1, \dots, n^D$ and $j = 1, \dots, m^D$. All potential sources are epidemiologically possible sources of infection, based on the times elapsed and the available biomarker data used to estimate the times elapsed. We use the term “potential sources” to describe epidemiologically possible sources, which we distinguish from “phylogenetically possible” sources, and “likely sources” following inference under the Bayesian model that we describe further below. The values of the $n^D \times m^D$ matrix $\boldsymbol{Y}^D$ are observed. Often, diagnosis times are used in lieu of infection time estimates[25,40]; in many cases these data are already highly informative for estimating $\boldsymbol{Z}$, for example in the case of new influenza outbreaks in school settings[41]. In most settings however, such as the spread of HIV at city-level or nationally[17,25,39], the number of potential sources to any new diagnosed case is typically very large and thus not very informative on $\boldsymbol{Z}$.

It is often possible to narrow down the potential sources of new cases based on demographic and clinical data, for example:

- exclude pairs with a time elapsed larger than 16 years, since each individual is likely to show symptoms within 8 years of seroconversion[42];

- using mortality data, exclude pairs in which the potential source died before the estimated infection date of the recipient [17];
- using migration and mobility data, exclude pairs in which the potential source did not reside in the study population by the estimated infection date of the recipient [43,44];
- using clinical data, exclude pairs with evidence that the potential source was not infectious on the infection date of the recipient. In our HIV case study, we estimated subject-specific viral load curves with LOESS smoothers to longitudinally collected viral load measurements and considered individuals with viral loads below 200 copies of virus per millilitre blood as non-infectious [45,46].

However, in our case study the number of potential sources for each new case continues to be very large with these exclusion criteria applied, which motivates us to consider pathogen sequence data.

A smaller subset of infected individuals also have a pathogen sequence sampled, and we denote the number of diagnosed individuals with a sampled pathogen by $n^S$ and the number of diagnosed new cases with a sampled pathogen in time period $\mathcal{T}$ by $m^S$. In the Netherlands, all newly diagnosed patients should have a partial HIV *polymerase* (*pol*) sequence sampled to determine potential drug resistance and suitable combination antiretroviral treatment for therapy [47]. Using population-level pathogen data, we construct HIV phylogenetic trees and perform ancestral state reconstruction with additional background sequences from outside of the study population to identify phylogenetically likely transmission chains circulating in the study population [39]. We reconstructed maximum-likelihood phylogenies with *FastTree* [48], which runs highly efficiently for a large number of taxa, and reconstructed ancestral states with *phyloscanner* [10], but other approaches could also be used [49–53]. Given the large molecular genetic diversity of HIV, these steps are performed separately for each of the predominant HIV subtypes and cirulating recombinant forms in the population [39,54]. Phylogenetically likely transmission chains are defined as groups of connected tips and internal nodes with corresponding ancestral states. In the case of HIV, most phylogenetically observed transmission chains are typically of size one, with no evidence of onward transmission in the study population. We focus only on those phylogenetically likely transmission chains with at least two members, defined by $\mathcal{C} = (C_1, \cdots, C_{n^\mathcal{C}})$, each containing a vector of vertices corresponding to its members. We denote a "phylogenetically possible" source of a new case by

$$Y_{ij}^S = \begin{cases} 1 \text{ if } \hat{T}_i < \hat{T}_j \cap \{i,j\} \in C_g \text{ for any } g = 1, \ldots, n^{\mathcal{C}} \\ 0 \text{ otherwise,} \end{cases} \tag{3}$$

where $i = 1, \ldots, n^S$ and $j = 1, \ldots, m^S$. The values of the $n^S \times m^S$ matrix $\boldsymbol{Y}^S$ are observed. For ease of reference we denote the phylogenetically possible transmission pairs by $\mathcal{P} = \{i = 1, \ldots, n^S, j = 1, \ldots, m^S | Y_{ij}^S = 1\}$, the phylogenetically possible sources of $j$ by $\mathcal{P}_j = \{i = 1, \ldots, n^S | Y_{ij}^S = 1\}$, and the total number of phylogenetically possible transmission pairs by $n^{\mathcal{P}}$.

## 2.3 Accounting for time elapsed when interpreting the patristic distance of virus from two individuals.

It is common to base phylogenetic inference of linkage on the number $D_{ij}$ of nucleotide mutations between pathogen sequences of $i$ and $j$ that occur along the shortest path in an estimated pathogen phylogeny. The patristic distance is non-negative and can be calculated with the R package `adephylo` [55] for all $i, j = 1, \ldots, n^S$, but here we restrict attention to $i \in \mathcal{P}_j$ and $j = 1, \ldots, m^S$. To account for differences in times from infection to sampling, we argue in this paper that inferences should also account for the time elapsed between pathogen samples since their divergence. To this end, we denote the sequence sampling dates of a sampled individual $i$ by $S_i$, where $S_i > \hat{T}_i$. To calculate the time elapsed, we make the approximation that the pathogen lineages leading to the observed samples in $i$ and $j$ diverged at the transmission event. Then, in the case that the sampling date of the source was before the estimated infection date of the recipient, we define time elapsed, $T^e_{ij}$ by

$$T^e_{ij} = (\hat{T}_j - S_i) + (S_j - \hat{T}_j). \tag{4}$$

In case the sampling date of the source was after the estimated infection date of the recipient, as illustrated in Figure 1, we have

$$T^e_{ij} = (S_i - \hat{T}_j) + (S_j - \hat{T}_j). \tag{5}$$

So, both cases can be subsumed under

$$T^e_{ij} = |S_i - \hat{T}_j| + (S_j - \hat{T}_j) \tag{6}$$

for all $i, j = 1, \ldots, n^S$, but again we will restrict attention to $i \in \mathcal{P}_j$ and $j = 1, \ldots, m^S$. Pathogens mutate over time, and so we expect that including the time elapsed into inferences will improve estimation of $\boldsymbol{Z}$, especially for rapidly evolving pathogens like HIV [56]. The data for inferring transmission flows in the study population are the patristic distances and estimated times elapsed across phylogenetically possible sources of new cases,

$$\boldsymbol{X} = \{D_{ij}, T^e_{ij} \mid (i,j) \in \mathcal{P}\}. \tag{7}$$

## 2.4 Signal component for the mixture model

We next develop a likelihood model for patristic distances and times elapsed from data of known transmission pairs, which will serve as the signal component of the BMM. Detailed knowledge of transmission chains is rare. However, one clinical investigation [57] in Belgium previously led to the characterization of the direction of transmission between individuals of one HIV transmission chain and the timing of transmission events. Subsequently, a large number of viral sequences were obtained for in-depth molecular analysis of this chain [58], and we developed the signal component of the BMM based on these data within a Bayesian random effects modelling framework due to the extensive variation in within-host viral evolution in these data [59]. In the Belgian study, HIV *pol* sequences were sampled for each individual in the known transmission chain from multiple time points. We therefore define each sequence pair for a source $i$ and recipient $j$ by $k = 1, \ldots, K_{ij}$ where $K_{ij}$ is the total number of sequence pairs available for each transmission pair $ij$. We also denote the entire Belgian transmission chain data with $\boldsymbol{B}$. Our resulting Bayesian hierarchical evolutionary clock model is

$$D_{ijk} \sim \text{Gamma}(\alpha_{ijk}, \beta_{ij}) \tag{8a}$$
$$\alpha_{ijk} = \mu_{ijk}\beta_{ij} \tag{8b}$$
$$\mu_{ijk} = (\gamma + \gamma_{ij})T^e_{ijk} \tag{8c}$$
$$\beta^{-1}_{ij} = \phi + \phi_{ij}, \tag{8d}$$

where $D_{ijk}$ is the $k^{\text{th}}$ patristic distance and $T^e_{ijk}$ is the $k^{\text{th}}$ time elapsed for pair $ij$. The Gamma observation likelihood is in shape-scale parameterisation such that its mean equals $\mu_{ijk} = \frac{\alpha_{ijk}}{\beta_{ij}}$ and its variance equals $\frac{\alpha_{ijk}}{\beta_{ij}^2}$, and the prior densities are

$$\log\gamma \sim \mathcal{N}(\log(10^{-2.5}), 0.2^2) \tag{9a}$$
$$\log\gamma_{ij} \sim \mathcal{N}(0, \sigma_\gamma^2) \tag{9b}$$
$$\sigma_\gamma \sim \text{Exp}(10) \tag{9c}$$
$$\log\phi \sim \mathcal{N}(0, 5^2) \tag{9d}$$
$$\log\phi_{ij} \sim \mathcal{N}(0, \sigma_\phi^2) \tag{9e}$$
$$\sigma_\phi \sim \text{Exp}(10). \tag{9f}$$

Here, $\mu_{ijk}$ denotes the expected evolutionary distance in the $k^{\text{th}}$ sequence pair for source $i$ and recipient $j$ and given the time elapsed $T^e_{ijk}$, where $\gamma$ corresponds to the overall log mean evolutionary rate across pairs and is given an informative prior [60], and $\gamma_{ij}$ are zero-mean random effects specific to transmission pair $ij$. The parameter $\phi$ corresponds to the mean degree of dispersion across pairs, while $\phi_{ij}$ are pair-specific zero-mean random effects. The model for the signal component may be adapted for other applications, for example if there is a consensus of evidence supporting non-linear evolutionary rates over time (see Supplementary Material S3).

The model in (8)-(9) was fitted with `cmdstanr` v.2.28.1 with 4 chains of 2000 samples each, including a burn-in of 500, and fitted the data well, as shown by Figure 2.

The predictive distribution of the shape and scale parameters of the model at given time elapsed for a new pair not in the training dataset, $T^e_{ij}$, is given by

$$\begin{gathered} f(\alpha^*_{ij}(T^e_{ij}), \beta^*_{ij}|\boldsymbol{B}) = \\ \int p(\alpha^*_{ij}(T^e_{ij}), \beta^*_{ij}|\gamma^*_{ij}, \phi^*_{ij})\, p(\gamma^*_{ij}, \phi^*_{ij}|\gamma, \phi, \sigma_\gamma, \sigma_\phi) \\ p(\gamma, \phi, \sigma_\gamma, \sigma_\phi|\boldsymbol{B})\, d(\gamma^*_{ij}, \phi^*_{ij}). \end{gathered} \tag{10}$$

We found that the predictive distribution obtained by replacing the posterior distributions for the hyper-parameters with their posterior medians was almost identical to (10) (Supplementary Figure S1), and for computational simplicity we used the latter.

## 2.5 Two component mixture model to identify transmission flows.

We next incorporate the fitted model (8)-(9) as a signal component for phylogenetically likely transmission pairs into a two component BMM. For ease of read we will from now on suppress in the notation that the signal component was derived on the data $\boldsymbol{B}$. We emphasize as others [61] that it is not possible to prove transmission between two individuals from the phylogenetic data, and we will throughout use the BMM transmission status allocations to each observation only as latent variables from which we deduce population-level patterns of transmission dynamics. The mixture model is constructed as follows. Conditional on the (unknown) transmission event having occurred ($Z_{ij} = 1$), we model the observed patristic distances for $(i, j) \in \mathcal{P}$ through

$$\begin{aligned} & p(D_{ij}|T^e_{ij}, Z_{ij} = 1) \\ = & \int \text{Gamma}(D_{ij}|\alpha^*(T^e_{ij}), \beta^*)\, f(\alpha^*(T^e_{ij}), \beta^*)\, d(\alpha^*(T^e_{ij}), \beta^*), \end{aligned} \tag{11}$$

where $f(\alpha^*(T^e_{ij}), \beta^*)$ is the posterior predictive distribution (10) at the estimated time elapsed $T^e_{ij}$. If there was no transmission event between individuals $i$ and $j$, we model

$$p(D_{ij}|T^e_{ij}, Z_{ij} = 0) = \text{Uniform}(0, d^{\max}), \tag{12}$$

where $d^{\max}$ is a suitably chosen large value, here $d^{\max} = 0.2$. We assume that the background distribution does not depend on time elapsed, but other choices are possible. The resulting BMM is

$$p(D_{ij}|T^e_{ij}) = \omega\, p(D_{ij}|T^e_{ij}, Z_{ij} = 1) + (1-\omega)\, p(D_{ij}|Z_{ij} = 0) \tag{13a}$$

$$logit(\omega) \sim \mathcal{N}(0, 4) \tag{13b}$$

for all $(i, j) \in \mathcal{P}$. Model (13) integrates out the latent transmission status variables $\boldsymbol{Z}$ and implicitly assumes that $Z_{ij} \sim \text{Bernoulli}(\omega)$, where $\omega$ is the mixing weight [34]. Since $\omega$ is assumed to be the same across all possible transmission pairs, we refer to (13) as the "vanilla mixture model", indicating it has no additional data informing $\omega$. Since $\omega$ is likely to decrease as the number of pairs under consideration increases, the variance for the prior on $\omega$ should be chosen to ensure support over the smallest value it could plausibly take, calculated as the number of incident cases divided by the total possible pairs without any exclusion criteria, $N/N(N-1)$. Note that the hyper-parameters of the signal component are specified through the data from the known Belgian transmission chain and the background component has no parameters, so the only parameters to be inferred are the mixing weight and the signal component random effects.

## 2.6 Estimating transmission flows from the latent, likely transmission pairs.

The primary purpose of the fitted BMM is to characterize population-level transmission flows between groups of individuals specified by distinct demographic, behavioral, or clinical covariates. We achieve this by considering all pairwise combinations of sampled individuals with source and recipient attributes $a$ and $b$ according to their posterior probabilities of being a transmission pair. There are two cases to consider and we begin by assuming that each sampled recipient $j$ has exactly one sampled possible source case $i$. Following notation in (1), the posterior transmission flows from group $a$ to group $b$ under the mixture model are

$$Z_{ab}|\boldsymbol{X} = \sum_{i\in a}\sum_{j\in b} \rho_{ij}|\boldsymbol{X} \tag{14a}$$

$$\rho_{ij}|\boldsymbol{X} = \frac{\omega\, p(D_{ij}|T^e_{ij}, Z_{ij} = 1)}{\omega\, p(D_{ij}|T^e_{ij}, Z_{ij} = 1) + (1-\omega)\, p(D_{ij}|Z_{ij} = 0)} \tag{14b}$$

$$\omega \sim p(\omega|\boldsymbol{X}) \tag{14c}$$

for all $a, b \in \mathcal{A}$, where $\rho_{ij}$ denotes the posterior probability that pair $ij$ is a transmission pair, also called the component membership probability in more general applications, and $p(\omega|\boldsymbol{X})$ denotes the posterior distribution of the mixing weights. If a pair involving the only phylogenetically possible source and the incident individual is outside the signal component, then it is likely incompatible with being a true transmission pair. Therefore, even if there is only exactly one phylogenetically possible source, it is possible that the proposed approach infers from the data that they are unlikely to be the true transmission source

For recipient $j$, if the data contain more than one phylogenetically possible source $i_1, \ldots, i_{L_j}$ with $n^{\mathcal{P}}_j > 1$ total possible sources, then the posterior transmission probabilities (14b) are obtained by considering that transmission can have occurred from exactly one or none of the $n^{\mathcal{P}}_j$ phylogenetically possible sources. For this, we quantify the posterior probability of events of the form $\{Z_{(i_1,j)} = 0 \cap \ldots \cap Z_{(i_{u-1},j)} = 0 \cap Z_{(i_u,j)} = 1 \cap Z_{(i_{u+1},j)} = 0 \cap \ldots \cap Z_{(i_{n^{\mathcal{P}}_j},j)} = 0\ \}$ for $u = 1, \ldots, n^{\mathcal{P}}_j$,

obtaining

$$
\begin{aligned}
&\rho_{i_u,j}|\boldsymbol{X} = \\
&\left(\omega\, p(D_{i_uj}|T^e_{i_uj}, Z_{i_uj}=1) \prod_{v\neq u} (1-\omega)\, p(D_{i_vj}|T^e_{i_vj}, Z_{i_vj}=0)\right) \Bigg/ \\
&\Bigg[\sum_{w=1}^{n^{\mathcal{P}}_j} \left(\omega\, p(D_{i_wj}|T^e_{i_wj}, Z_{i_wj}=1) \prod_{v\neq w} (1-\omega)\, p(D_{i_vj}|T^e_{i_vj}, Z_{i_vj}=0)\right) + \\
&\qquad \prod_{v=1}^{n^{\mathcal{P}}_j} (1-\omega)\, p(D_{i_vj}|T^e_{i_vj}, Z_{i_vj}=0)\Bigg]
\end{aligned} \tag{15}
$$

for all $u = 1, \ldots, n^{\mathcal{P}}_j$ and then evaluate $Z_{ab}|\boldsymbol{X} = \sum_{i_u\in a}\sum_{j\in b}\sum_{u=1}^{n^{\mathcal{P}}_j} \rho_{i_uj}|\boldsymbol{X}$.

## 2.7 Incorporating additional covariates.

We next considered incorporating additional individual-level covariates as a predictor on the BMM mixing weights to better separate true transmission pairs from those with false transmission pair signal, i.e. truly unlinked pairs of individuals that fall into the signal component of the mixture model (11). We denote $p$-dimensional covariates for all $n^{\mathcal{P}}$ phylogenetically possible pairs with $\boldsymbol{C} \in R^{n^{\mathcal{P}}\times p}$. These could be the age of each individual in a pair, socio-demographic characteristics, place of birth, or even pairwise properties such as being part of the same household. This updates the vanilla mixture model (13) to the "covariate" mixture model

$$
\begin{aligned}
p(D_{ij}|T^e_{ij}) &= \omega_{ij}\, p(D_{ij}|T^e_{ij}, Z_{ij}=1) + (1-\omega_{ij})\, p(D_{ij}|Z_{ij}=0) && \text{(16a)}\\
logit(\omega_{ij}) &= \eta_0 + \boldsymbol{C}_{ij}\boldsymbol{\eta} && \text{(16b)}\\
\eta_0 &\sim \mathcal{N}(0,4) && \text{(16c)}\\
\boldsymbol{\eta} &\sim \mathcal{N}(0,1), && \text{(16d)}
\end{aligned}
$$

where $C_{ij}$ is the row of the matrix of covariates corresponding to the $ij^{\text{th}}$ pair and $\eta_0 \in R$ and $\boldsymbol{\eta} \in R^p$ are additional model parameters.

In many cases the covariates can be continuous, such as age of the likely source and the likely recipient individual in the $ij^{\text{th}}$ pair. In these settings, we model the dependency of the mixing weights on the covariates through univariate or bivariate random functions. In the latter case, for computational efficiency, we use zero-mean two-dimensional Hilbert-space Gaussian Process (HSGP) approximations, which scale more efficiently with a large number of observations compared to bivariate Gaussian Process priors[29,62,63]. This updates the vanilla mixture model to the "HSGP random function" mixture model

$$
\begin{aligned}
p(D_{ij}|T^e_{ij}) &= \omega_{ij}\, p(D_{ij}|T^e_{ij}, Z_{ij}=1) + (1-\omega_{ij})\, p(D_{ij}|Z_{ij}=0) && \text{(17a)}\\
logit(\omega_{ij}) &= \eta_0 + \boldsymbol{f}(x_{ij,1}, x_{ij,2}) && \text{(17b)}\\
\boldsymbol{f} &\sim \text{HSGP}(0, k((x_1,x_2),(x_1,x_2)')) && \text{(17c)}\\
k((x_1,x_2),(x_1,x_2)') &= \alpha \exp\left(\frac{(x_1-x_1')^2}{2\ell^2_{x_1}} + \frac{(x_2-x_2')^2}{2\ell^2_{x_2}}\right) && \text{(17d)}\\
\eta_0 &\sim \mathcal{N}(0,4) && \text{(17e)}\\
\alpha &\sim \mathcal{N}(0, 0.15^2) && \text{(17f)}\\
\ell_{x_1}, \ell_{x_2} &\sim \text{Inv-Gamma}(5,5), && \text{(17g)}
\end{aligned}
$$

where $\alpha \in R$, and $\ell_{x_1}, \ell_{x_2} \in R^+$ are additional model parameters. Further details on the HSGP regularising prior density are in the Supplementary Material S2. Transmission flows are calculated from the posterior as before (15).

## 2.8 Numerical inference.

Throughout, the model was fitted with the Stan probabilistic computing language using the `cmdstanr` interface, v2.28.1, with 4 chains of 2000 samples each, and a burn-in of 500. Code is available at `github.com/MLGlobalHealth/source.attr.with.infection.time`. There were no observed divergences, and the minimum number of effective samples across all parameters was 2065 on the simulated data and 3492 on our application to Amsterdam data; see also the Supplementary Material S1 for diagnostic plots.

## 2.9 Simulations to evaluate estimation performance.

We assessed the performance of the mixture models for estimating transmission flows and transmission sources on simulated HIV transmission networks derived from the discrete-time individual-based HIV epidemic model (PopART-IBM), which was developed contextually to the HPTN071/PopART HIV combination intervention prevention trial[64]. The model is informed by data collected from surveys, health care facilities, and by community health care workers delivering interventions as part of the HPTN071 trial. The model parameters were calibrated to age-sex-specific data on incidence, prevalence, ART uptake, and viral suppression from a representative cohort at the level of communities participating in the trial. Simulated transmission events depend on individual-level parameters including age, sex, set point viral load, and CD4 counts of the transmitter. The simulation begins with a starting population size in 1900, and the epidemic is seeded randomly with infectious cases in 1965-1970. The model runs until 2020 and returns a large number of separate, simulated transmission chains. For each new case, sequence sampling dates were simulated using a Weibull distribution. For each transmission pair, the time elapsed was calculated and patristic distances were simulated from the signal distribution (8). For unlinked pairs, patristic distances were simulated from the two-dimensional Uniform background distribution (12). Full details are presented in the Supplementary Material S4.

## 2.10 Evaluating estimation performance.

For each of the models (13), (16) and (17), the accuracy of the target quantities (1) from the posterior was evaluated by comparing the mean absolute error (MAE), defined for each MC sample, to the corresponding true quantity in the simulated data among known transmission pairs. For (1a), $\text{MAE} = \sum_{a,b \in \mathcal{A}} |\pi_{ab} - \pi^*_{ab}| / |\mathcal{A}|$, where $\pi^*_{a,b}$ are the true simulated flows from group $a$ to group $b$. The posterior MAE was summarized by taking the median and 95% quantiles. The MAE for the remaining target quantities were evaluated analogously.

# 3 Results

## 3.1 Constructing the signal component of the Bayesian mixture model

We first fitted the Bayesian hierarchical evolutionary clock model (8)-(9) to patristic distances and time elapsed from 5,186 sequences from ten individuals in a well-characterized HIV transmission chain from Belgium[57,58]. Maximum-likelihood viral phylogenies were reconstructed with `RAxML` v7.4.2[65]. Excluding pairs involving one multi-drug resistant individual, patristic distances between 2,807 sequence combinations from eight individuals involved in seven known transmission events were computed from the reconstructed phylogeny. Infection times were ascertained to narrow time ranges of up to a few months in the study, and we used the midpoint of these uncertainty ranges. The time elapsed was calculated according to (6) for each combination of viral sequences from the recipient and transmitting partner of each transmission event. Figure 2A illustrates the patristic distances relative to time elapsed for all 2,807 pairwise sequence combinations, with data corresponding to each of the seven transmission events shown in a different colour. We found substantial heterogeneity in the rate of evolution across transmission pairs, prompting us to use the Bayesian hierarchical model (8)-(9) to estimate an overall mean evolutionary clock and associated uncertainty range. The model fit the data well, with 96% of the observed genetic distances within 95% posterior predictive credible intervals (Figure 2B). The estimated overall mean evolutionary rate was $4.5 \times 10^{-3}$ [95% CrI $3.5 \times 10^{-3}$ - $5.6 \times 10^{-3}$] substitutions per site per year, compatible with previously published estimates[58]. Figure 2A shows in orange the shape of the estimated posterior predictive distribution of patristic distances under the model for each 0.1 year increment in time elapsed. These posterior predictive distributions define the signal component of the BMM for phylogenetic source attribution, and for reasons of parsimony we chose a uniform distribution as the background component (see Methods).

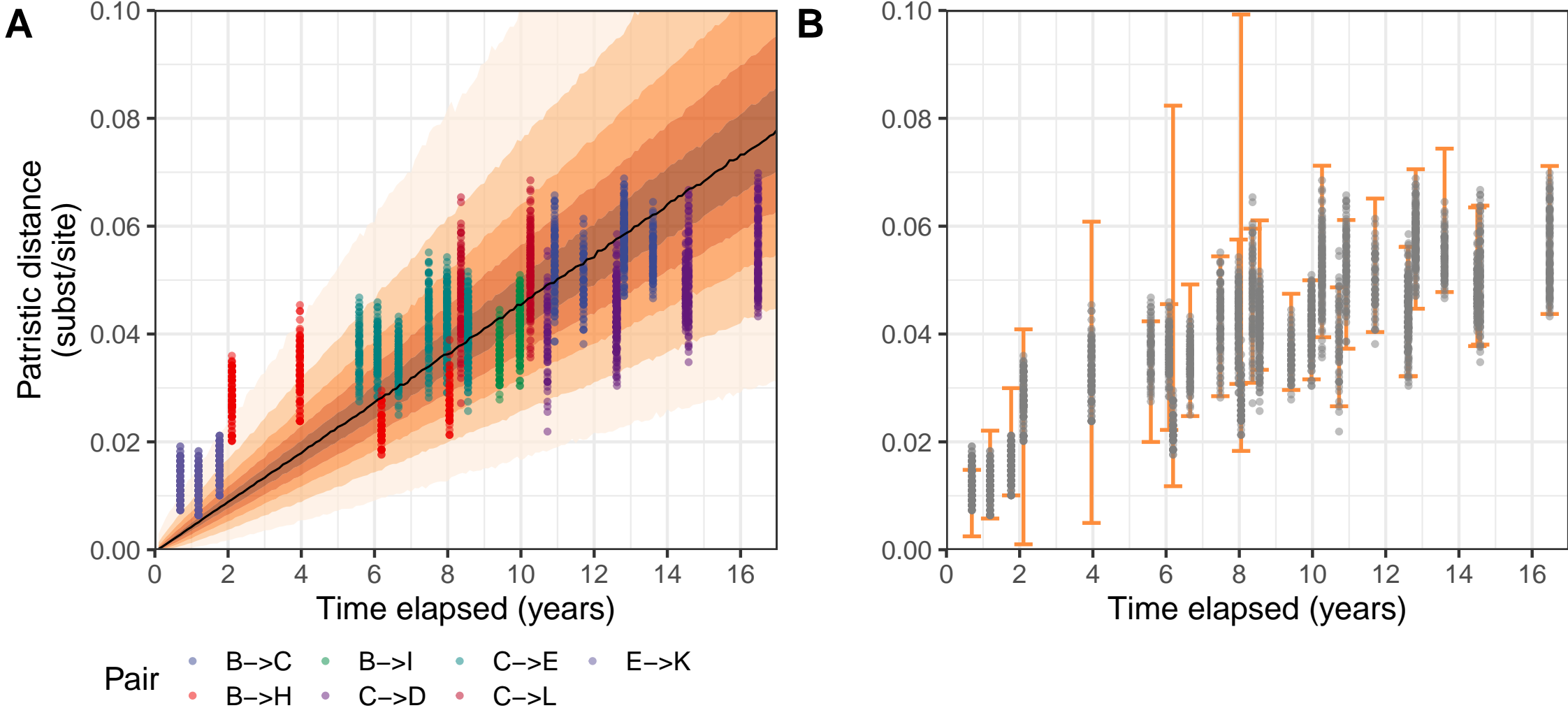

Figure 2: **Belgian transmission chain data and model fit of evolutionary clock model.** (**A**) Patristic distances between pairs of sequences from epidemiologically confirmed transmission pairs in Belgium, against time elapsed. Colours denote transmission pairs. Orange ribbons correspond to quantiles (at 10% increments between 10% and 90%, in addition to 95% credible intervals) of the posterior predictive distribution of the of the average evolutionary rate across pairs in the hierarchical clock model. (**B**) Posterior predictive 95% credible intervals of patristic distances by time elapsed (orange) for known transmission pairs, with data points overlaid. Transmission pair B→H leads to large uncertainty intervals at times elapsed 2-8 but are retained in the model, which allows for between-pair heterogeneity, to increase the number of training data points.

## 3.2 False transmission pair signal in the BMM signal component

We next sought to assess the estimation accuracy of the two-component vanilla BMM (13) on simulated data generated under an individual-based HIV epidemic model developed to represent contemporary African HIV epidemics and used to interpret the outcomes of the HIV prevention trial network trial 071 (HPTN071/PopART)[64]. The model simulated transmission dynamics over 55 years from 1965 in a starting population of 32,217, initialising the epidemic with 148 transmission events to randomly chosen individuals in the first five years under default input parameters (see Supplementary Material S4). The model simulated 34,961 transmission events. For ease of tractability, we considered only the most recent 500 incident cases between 09/06/2017 and 31/12/2019, which were distributed over 276 distinct transmission chains (Figure 3A). We assumed that a pathogen sequence was sampled for all individuals with HIV.

Since 2005, there were 9,355 potential sources with infection dates that preceeded those of a new case between mid 2017 and end of 2019. These formed 4,552,006 unique pairwise combinations of potential heterosexual transmission pairs with the 500 incident cases. 768,257 potential pairs were excluded based on very large time elapsed, including 39 actual transmission events, due to the epidemic model simulating individual infection times which can lead to unrealistic values from the tails of the distribution in some cases (see Methods). The 461 actual transmission events thus comprised 0.012% among all potential transmission pairs, rendering the unlabelled classification task of identifying true transmission pairs highly imbalanced.

Individuals were grouped into distinct phylogenetic transmission networks, reducing the number of heterosexual, phylogenetically possible transmission pairs to 5,519, with the 461 actual transmission events comprising 8.3% of the remaining pairs, a 600-fold improvement in balance. Each newly acquired case had on average 9,104 potential sources and 11.2 phylogenetically possible sources after excluding potential sources not within the same transmission chain (Figure 3B). Figure 3C illustrates how the quantitative data on patristic distances and time elapsed narrow down further the likely transmitters among pairs that fall into the signal component of the BMM. Importantly, a considerable fraction of the transmission events in the simulation (shown in red in Figure 3C), and their true sources, were associated with late diagnosis, which resulted in a large time elapsed. Application of standard phylogenetic selection criteria for reconstructing transmission pairs, such as a maximum patristic distance of 1.5% (i.e. 1.5 substitutions per 100 nucleotides [66–69], indicated by the horizontal line in Figure 3C), would have excluded these pairs, introducing selection bias. By accounting for time elapsed in the phylogenetic source attribution problem, we avoid this selection bias. However, the inference problem remains challenging, since 37% of phylogenetically possible pairs exhibited false transmission pair signal, defined as truly unlinked pairs that fall into the 95% quantiles of patristic distances for given time elapsed under the BMM signal component.

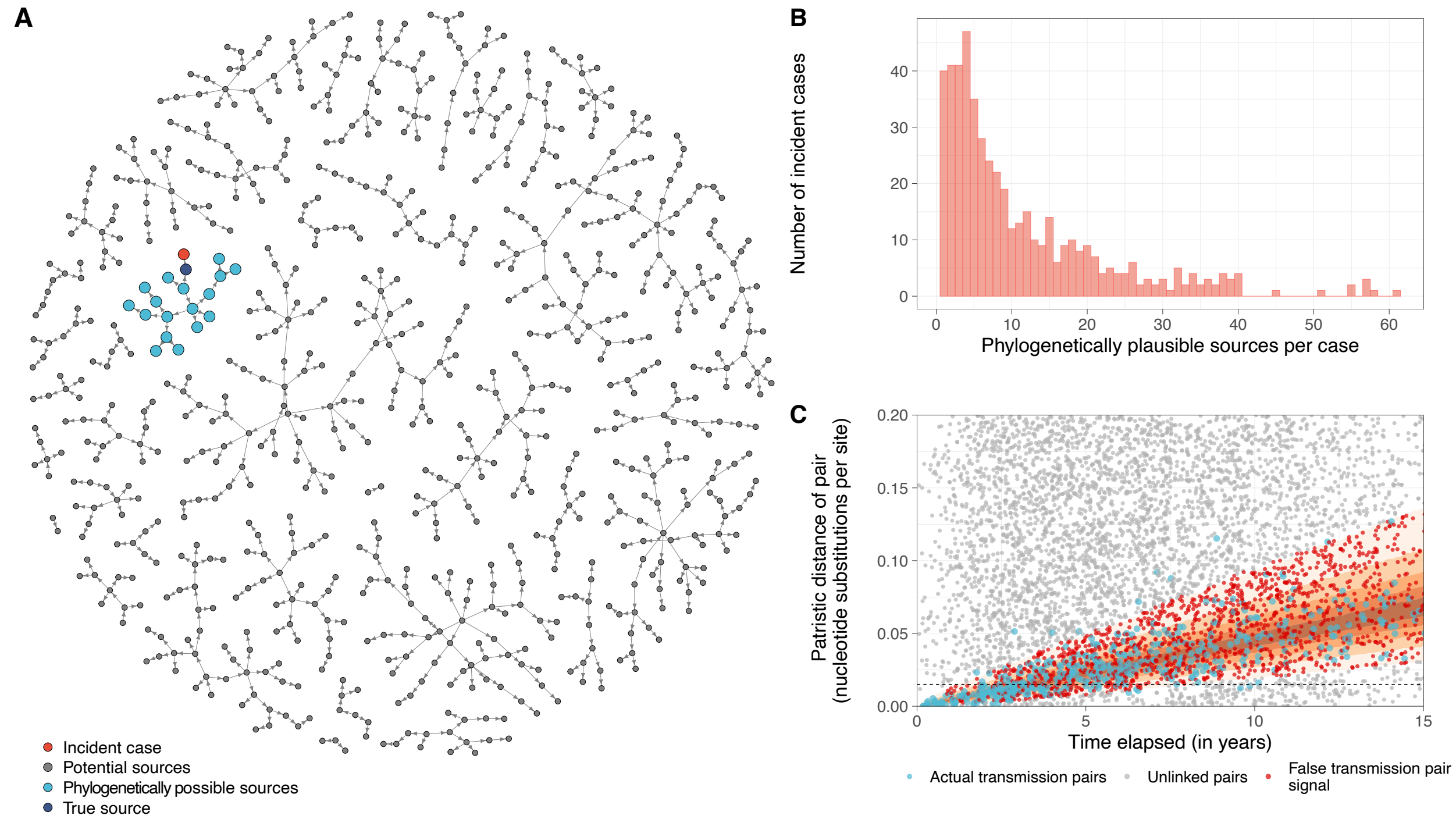

Figure 3: **Phylogenetically possible sources and false transmission pair signal in an HIV epidemic simulation.** (**A**) Sample of transmission chains containing one of the 500 most recent incident cases in the simulation generated under the PopART-IBM. One of the recent cases is highlighted in red, a random sample of their potential sources in grey, which correspond to sampled participants who are part of a distinct phylogenetically observed transmission chain, all phylogenetically possible sources in light blue, and the actual transmitting partner of the simulation in dark blue. (**B**) Number of all phylogenetically possible sources for the 500 most recent incident cases in the simulation. (**C**) Patristic distances and time elapsed between phylogenetically possible sources and each of the 500 most recent incident cases in the simulation. Actual transmission pairs in the simulation are shown in red. Pairs that exhibited false transmission pair signal are shown in light green. These are defined as phylogenetically possible unlinked pairs with patristic distances that fall within the 95% quantiles of the posterior predictive distribution of the distances for a given time elapsed from the fitted molecular clock model. All other unlinked pairs are in grey.

### 3.3 Accuracy in phylogenetic source attribution with the Bayesian mixture model with little false transmission pair signal

With these statistical challenges in mind, we first evaluated the estimation accuracy of the BMM in the idealized scenario that the newly acquired cases had each on average two phylogenetically possible sources, including the true transmitting partner. The unlinked pairs were sampled randomly to achieve the desired balance. On these pairs, we considered the following binary source attribution problem. In each of the 461 transmission events, the true transmitting partners were allocated to have the same population-level characteristic "category 1" while all other phylogenetically possible but unlinked sources were allocated to "category 2". In this simulation, all phylogenetically possible pairs had small patristic distance ($< 0.2$) (Figure 4). Of these, 13% of unlinked pairs had false transmission pair signal (i.e. lying within the 95% quantiles of the BMM signal component), and 1.5% of true transmission pairs had no transmission pair signal (i.e. lying outside the 95% quantiles of the signal component) (Figures 4B). We then fitted the vanilla BMM (13) to these data, aiming to estimate that 100% of transmissions originated from category 1. Figure 4C illustrates the posterior probabilities that each pair are classified by the BMM as a true transmission pair. A small number of true transmission pairs exhibiting no signal, were inferred by the model to be likely transmission pairs (posterior transmission pair probability $> 0.5$), likely the result of lying close to the edges of the posterior predictive distribution of the signal component. Weighting each phylogenetically possible pair by the posterior transmission pair probabilities, an estimated 89% (88-90%) of transmissions originated from category 1 (Figure 4D), implying a MAE for the sources estimated among the entire population (1c) of 11% [10-12%]). For comparison, if pairs were classified as phylogenetically likely transmission pairs as commonly done based on patristic distances less than 1.5%[66–69], 73% of transmissions were inferred to originate from category 1, implying a MAE of 27%.

### 3.4 Accuracy in phylogenetic source attribution with the Bayesian mixture model with substantial false transmission pair signal

We next returned to simulated transmission dynamics under the HPTN071/PopART model that included many more pairs with false transmission pair signal in the BMM signal component. We considered the same binary source attribution problem as before, allocating all actual transmitting partners to have the same population-level characteristic "category 1" and all unlinked, but phylogenetically possible sources to "category 2". In this simulation, 26% of pairs were unlinked with false transmission pair signal, and 0.3% of pairs had no transmission pair signal despite being linked transmission pairs (Figure 5B). Fitting the vanilla BMM (13), we estimated that 45% (43-47%) of transmissions were attributed to category 1, implying a MAE of 55% (52-57%). For comparison, when we classified phylogenetically likely transmission pairs based on patristic distances less than 1.5%, we estimated that 25% of transmissions were attributed to category 1, and the MAE was 75%. These simulation results indicate that the error in phylogenetic source attribution can be very large with either a standard classification approach or the vanilla BMM. In particular, the large estimation error is associated with situations in which the phylogenetically possible pairs that are falsely classified as transmission pairs outnumber those pairs that are correctly classified.

To improve on this poor estimation accuracy, we next considered the covariate BMM (16) that included as predictor to the BMM mixing weights (16c) the binary covariates: "category 1" and "category 2". In other words, we added features that make the classification task perfectly identifiable upon correct parameterisation of the BMM. Fitting the model, we found that nearly all previous pairs with false transmission pair signal were now estimated to have a low posterior probability of being a transmission pair (Figure 5C) and 89% [82-97%] of transmission events were attributed to category 1, implying a MAE of 11% (3-18%) (Figure 5D). These simulations indicate that the covariate BMM can accurately estimate population-level drivers of transmission even when many pairs exhibit false transmission pair signal, provided that additional covariate data is available that perfectly separates transmission pairs from unlinked pairs.

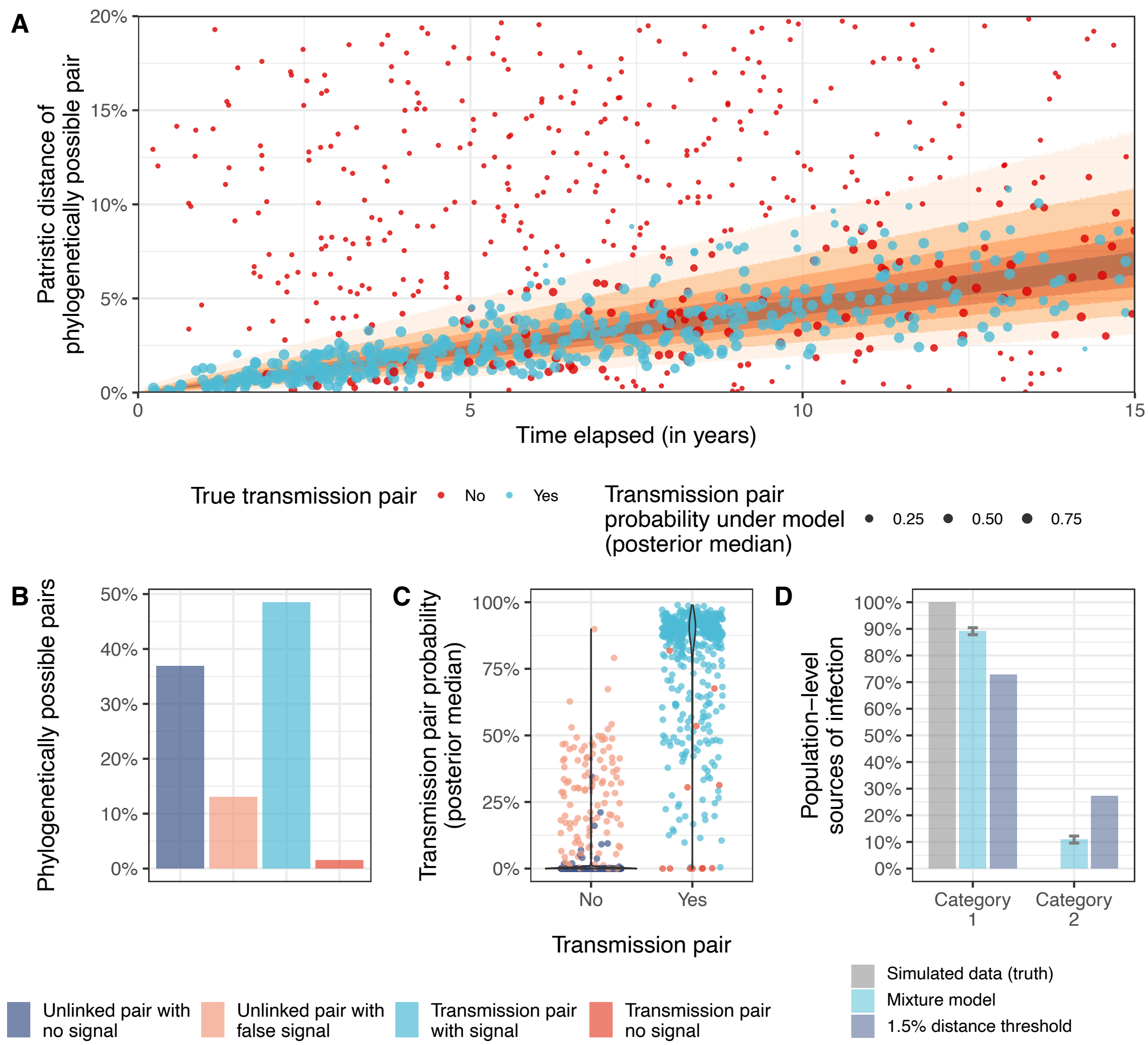

Figure 4: **Source attribution with the vanilla BMM in the case of two competing phylogenetically possible sources per new case.** From the HIV epidemic simulation, only two competing phylogenetically possible sources for each of the 500 most recent cases were retained to assess source attribution in an idealized scenario. (**A**) Patristic distances and time elapsed for each phylogenetically possible pair, coloured by their posterior probability of being a true transmission pair in the vanilla BMM. (**B**) Empirical proportion of pairs with and without transmission pair signal (based on 95% quantiles of molecular clock), for actual transmission pairs and unlinked pairs among the phylogenetically possible pairs. (**C**) Posterior probability of being a transmission pair, by actual transmission pairs and unlinked pairs. (**D**) Posterior median estimates of transmission flows and 95% credible intervals under the vanilla BMM, compared to using a 1.5% threshold on patristic distances, and simulated ground truth.

More realistically, we considered a (non-binary) source attribution scenario in which simulated transmission events were associated with similarity in the age of newly acquired cases and their

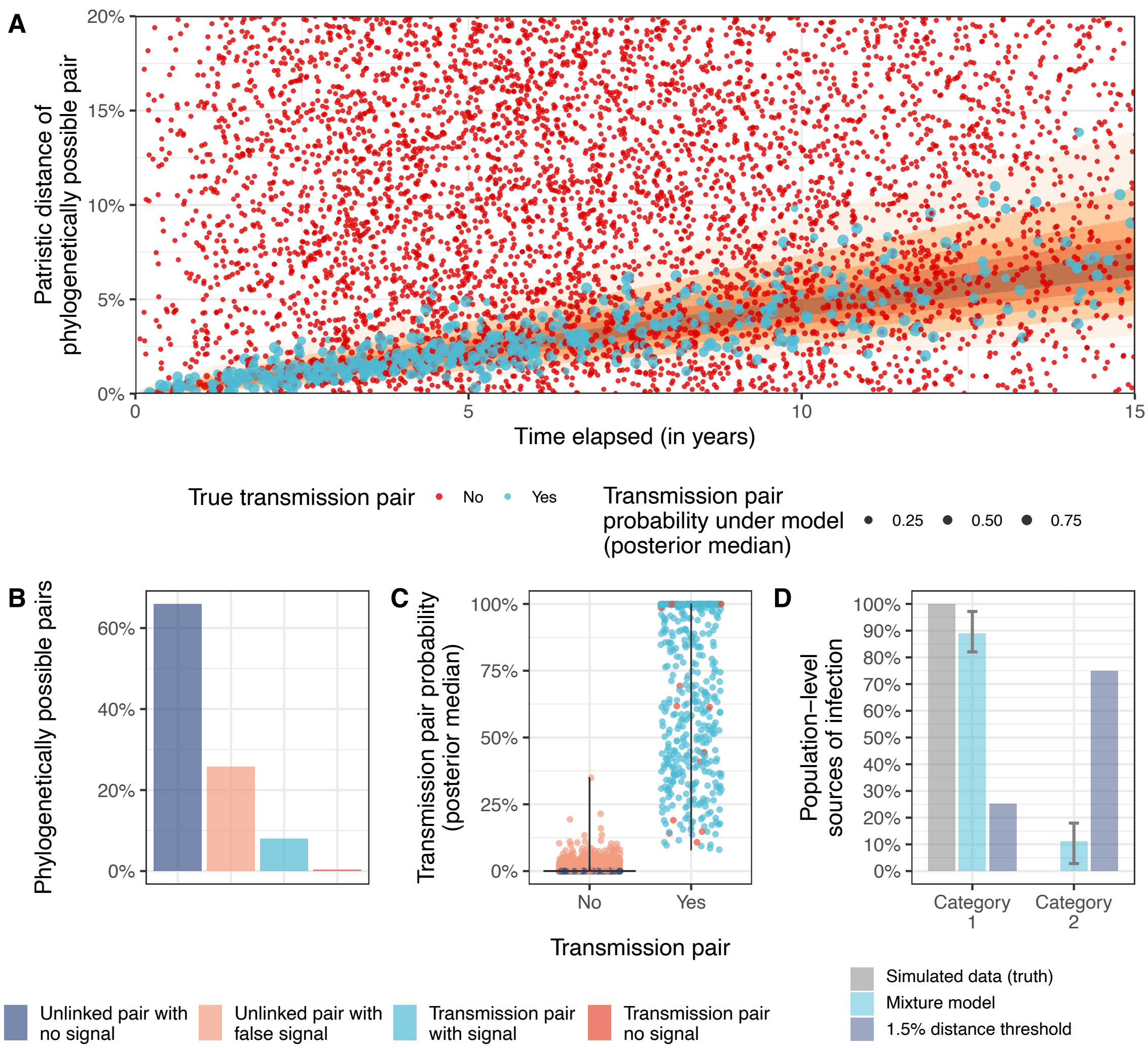

Figure 5: **Source attribution with the covariate BMM in the case of many competing phylogenetically possible sources per new case.**(**A**) Patristic distances (number of substitutions per 100 nucleotides) and time elapsed for each phylogenetically possible pair, coloured by their posterior probability of being a true transmission pair. (**B**) Empirical proportion of pairs with and without transmission pair signal (based on 95% quantiles of molecular clock), for actual transmission pairs and unlinked pairs. C) Posterior probability of being a transmission pair, split by actual transmission pairs and unlinked pairs. (**D**) Posterior median estimates of transmission flows and 95% credible intervals under the covariate BMM, compared to using a 1.5% threshold on patristic distances.

transmitting partners. Data were simulated so that age similarity did not perfectly separate transmission pairs from unlinked pairs (Figure 6A). For the actual transmission pairs, we simulated the age of the sources using a truncated log-normal distribution between 16 and 75 with a mean of 30 years, and simulated the age of the recipients in each pair using the age of their source as mean to generate correlated ages. We simulated the ages of the unlinked pairs uniformly across the same age range. We included the ages of the cases and their phylogenetically possible sources as

| | | Mean absolute error (%)* | | |
|---|---|---|---|---|
| Average number of sources per recipient | Pairs classified by patristic distances < 1.5% | Vanilla BMM | Covariate BMM† | HSGP random function BMM‡ |
| 2 | 2.3% | 0.9% [0.7-1.1%] | 0.5% [0.3-0.8%] | 0.3% [0.2-0.5%] |
| 2.5 | 2.6% | 1.3% [1.1-1.5%] | 0.5% [0.3-0.7%] | 0.3% [0.2-0.5%] |
| 3.3 | 3.7% | 1.9% [1.6-2.1%] | 0.9% [0.6-1.2%] | 0.6% [0.4-0.9%] |
| 5 | 4.2% | 2.9% [2.5-3.2%] | 0.7% [0.4-1.2%] | 0.7% [0.4-1.1%] |
| 10 | 5.8% | 3.7% [3.4-4.1%] | 1.4% [1-1.8%] | 1% [0.7-1.4%] |
| 12.5 | 5.2% | 4.2% [3.9-4.5%] | 1.8% [1.4-2.4%] | 1.2% [0.8-1.7%] |

* Mean absolute error reports the estimation error in the attributed source categories from each model.

† Covariate BMM includes categorical information on both the source and recipient at the time of infection of the recipient.

‡ HSGP BMM incorporates a continuous variable (e.g. 1 year ages) on both the source and recipient at the time of infection of the recipient.

Table 1: **Accuracy in phylogenetic source attribution by 5-year age bands on simulated data with substantial false transmission pair signal**.

independent predictors to the mixing weights of the covariate BMM, and sought to recover the age profile of the sources of transmission in the simulation by 5-year age groups. We performed several experiments, keeping in each experiment the average number of phylogenetically possible sources per incident cases fixed at 2, 2.5, 3.3, 5, 10, 12.5, to represent scenarios with minimal to pervasive false transmission pair signal (Figure 6B). We also compared source attribution using a 1.5% patristic distance threshold to classify phylogenetically likely transmission pairs, the vanilla BMM, and a HSGP random function BMM (17) in which the mixing weights were modelled through a 2D random function on the age of the newly acquired case and their phylogenetically possible source. Table 1 shows that the HSGP random function BMM had consistently lowest MAE, and was 0.3% [0.2-0.5%] for an average of two phylogenetically possible sources per newly acquired case, increasing to 0.7% [0.4-1.1%] for an average of five phylogenetically possible sources per newly acquired case, and 1.2% [0.8-1.7%] for an average of 12.5 phylogenetically possible sources per case 1. Note that the MAE cannot be directly compared with earlier results since we estimated the sources across more strata.

## 3.5 Age-specific drivers of transmission in Amsterdam MSM transmission chains

Finally, we illustrate our method on data from HIV transmission chains among Amsterdam MSM. Data on people living with HIV in the Netherlands were obtained from the ATHENA observational cohort, comprising of demographic, clinical and viral sequence data for diagnosed individuals until January 17, 2022[37]. Amsterdam residents were geolocated using residential postal codes at enrolment, or at a registration update. Longitudinal CD4 and viral load measurements enabled us to estimate HIV infection times with a Bayesian model trained on data of 19,788 seroconverters with known date of last negative test from the CASCADE collaboration[38,39]. We used posterior median infection time estimates to calculate the time elapsed (6). HIV phylogenies and phylogenetic transmission chains for the major subtypes and circulating recombinant forms (CRFs) among Amsterdam MSM were reconstructed in the context of 1,321 additional HIV sequences from Amsterdam residents in other HIV risk groups, 7,119 from the rest of the Netherlands and 12,821 international sequences using *FastTree* and *phyloscanner* (see Methods). Figure 7A illustrates the resulting data for a large clade of all subtype B samples.

Our primary aim was to quantify age-specific population-level sources of transmissions among

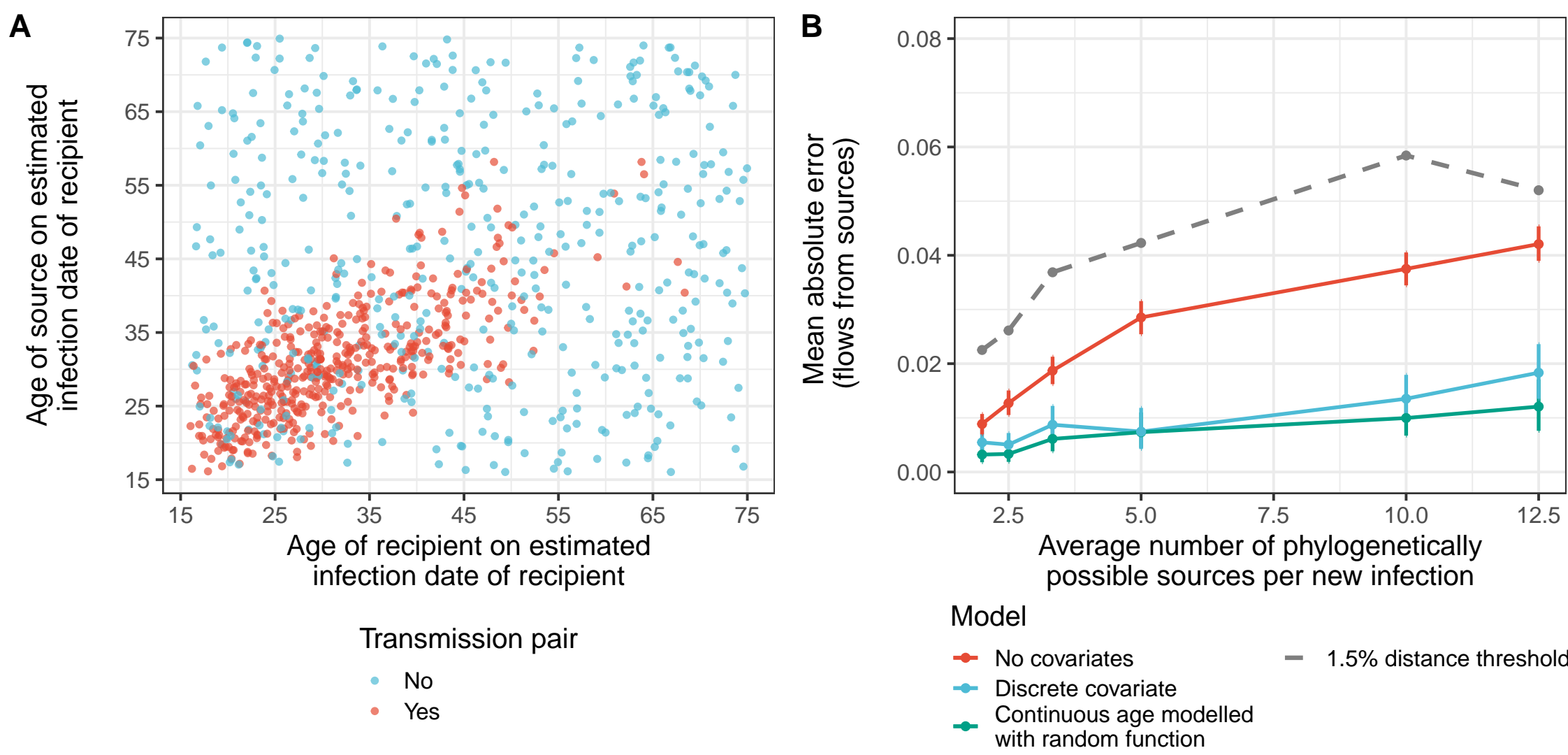

Figure 6: **Performance of three models, benchmarked against inference through a fixed genetic distance threshold.** A) Structure of the simulated ages of sources and recipients in true transmission pairs and unlinked pairs. B) Mean absolute error (MAE) in the estimated transmission flows from each source category (age group) under the three models. MAE for flows by estimating sources with a 1.5% patristic distance threshold. We note that the MAE cannot be directly compared to that in Figure 4 because there are more strata from which we estimate the sources.

Amsterdam MSM between 2010-2021. In total, there were 1,335 new HIV diagnoses among Amsterdam MSM with an estimated infection date during the study period, and 840 had an HIV sequence of one of the predominant HIV-1 subtypes or CRFs B, C, 01AE, A1, 02AG, D, G, F1, 06cpx available. Of these, 524 Amsterdam MSM formed a phylogeographically distinct transmission chain of more than one member, which we focus on in our analysis. The remainder were considered to have acquired HIV from a non-Amsterdam MSM source, or unobserved Amsterdam MSM. For these, we identified 3,033 potential sources among Amsterdam MSM with an estimated infection date that preceded that of the new case, that formed 1,372,332 potential transmission pairs. Of these, we excluded 65,104 (4.74%) pairs with potential sources who were deceased prior to the infection date, 16,934 (1.23%) with potential sources who migrated to the Netherlands after the infection date, 906,756 (66.07%) with potential sources who were estimated to have suppressed and thus untransmittable virus at the infection date, 13,140 (0.96%) with potential sources who had an implausibly long time elapsed, and 367,574 (26.78%) with potential sources who were not part of the same phylogenetically reconstructed transmission chain. Seven of the remaining phylogenetically possible pairs had a distance of zero (i.e. their sequences were identical). Examining metadata for these pairs suggested these were distinct individuals, so we reset their patristic distance to one mutation across the length of the alignment (0.077% substitution rate), but carried out a sensitivity analysis to assess the impact of omitting these pairs (See Supplementary Material S6.3).

In total, this left 409 sampled new Amsterdam MSM cases with at least one sampled Amsterdam MSM as phylogenetically possible source. Each new case had a median of 3 and on average of 6.9 phylogenetically possible sources, and there were a total of 2,824 phylogenetically possible transmission pairs. The median time elapsed was 4.17 years, and thus we expected many phylogenetically possible transmission pairs to have patristic distances above a patristic distance threshold of 1.5% since the dates at which viral sequences could be obtained were long after the estimated infection time. Many phylogenetically possible pairs had patristic distances outside of the posterior predictive distribution of the clock model, suggesting these are unlikely to be true transmission pairs given their incompatibility with the molecular clock (Figure 7B). A small proportion of pairs had small patristic distance relative to their time elapsed. These could be true transmission pairs, with individual-level infection date uncertainty resulting in error in the estimated time elapsed, or could be unlinked pairs from the same phylogenetic cluster, with an intermediary person between them in the unobserved transmission chain. We fitted the HSGP random function BMM (17), with the mixing weights specified through a 2D random function on the age of the recipient and the source at the estimated infection time of each phylogenetically possible transmission pair. The model converged and mixed well with no divergences and a runtime of 59 minutes on a 2020 MacBook Pro (Supplementary Figures S4-S5). This model had a a mean absolute error of $< 1.2\%$ in simulations configured to the same average number of pairs per incident case (Table S2), and out of all models considered, was the model with highest expected log posterior density (Supplementary Table S3). Only 2 phylogenetically possible transmission pairs were associated with posterior transmission probabilities above 95% but even for these we cannot rule out unsampled intermediates or unsampled sources, and therefore cannot interpret the corresponding phylogenetically possible sources with any certainty as the actual transmitting partner. We thus only considered the phylogenetically possible sources as weighted by their posterior transmission probabilities in aggregate, and focus on their population-level age characteristics.

Infection times were bootstrap sampled 50 times and models refitted to account for uncertainty in infection time estimates. We found that transmissions originated from all age groups among Amsterdam MSM in 2010-2021, with an estimated 28% [23-33%] from 15-29 year olds, 31% [27-36%] from 30-39 year olds, 26% [22-31%] from 40-49 year olds, and 15% [12-19%] from MSM aged 50 and above (Figure 8A). Stratifying by age of recipients, we found that most incident cases had a source within the same age band, but not strongly so. For all age groups, more than half of incident cases originated from sources that were either older or younger. For example, for Amsterdam MSM aged 15-29, an estimated 39% [32-46%] of incident cases originated from 15-29 year olds, 34% [28-41%] from 30-39 year olds, 19% [14-25%] from 40-49 year olds, and 7% [4-11%] from MSM aged 50 and above (Figure 8B). Considering transmissions by age of the recipients, our data indicated that the age structure of transmission sources is shifting from older sources to incident Amsterdam MSM aged 15-29 to younger sources to incident Amsterdam MSM aged 50 and above. To examine this further, we calculated the age gap between the age of the phylogenetically possible source and the age of the new case at the likely time of the infection event and weighted these age gaps by the posterior probability that the pair represents a transmission event (Supplementary Material S5.1). We estimate that 15-29 year old Amsterdam MSM had sources of transmissions who were on average 6 years older (posterior interquartile range IQR 0 to 15 years) while 30-39 year olds had sources on average the same age (IQR 6 years younger to 9 years older), 40-49 year olds had sources on average 6 years younger (IQR 14 years younger to 2 years older), and Amsterdam MSM aged over 50 had sources that were on average 11 years younger (IQR 3 to 20 years younger).

# 4 Discussion

We present a novel approach to phylogenetic source attribution that combines molecular genetic distances between sampled pathogen sequences with time elapsed in a Bayesian mixture model that has a population-level molecular clock as its signal component. This statistical approach to source attribution is now becoming increasingly possible as various methods are now available for estimating infection times from clinical and demographic data as we have used here[70,71], but also from deep-sequence data[72–75], the number of ambiguous nucleotide mutations in consensus sequences[76–79], and combinations of biomarker and pathogen sequence data[80,81]. The main ad-ditional information needed to leverage these time since infection estimates is the evolutionary clock model that underpins the signal component of the Bayesian mixture model. We focused on developing the approach for HIV transmission dynamics because a previous study[58] sequenced a large number of pathogens from a known transmission chain, enabling us to use a large data set of over 2,800 sequence pairs to construct the signal component. Similar approaches, or even using molecular clocks parameterized as part of large-scale phylogenetic analyses[82], could be explored for a wide range of other pathogens that mutate sufficiently rapidly relative to the time scale of transmission dynamics[83,84].

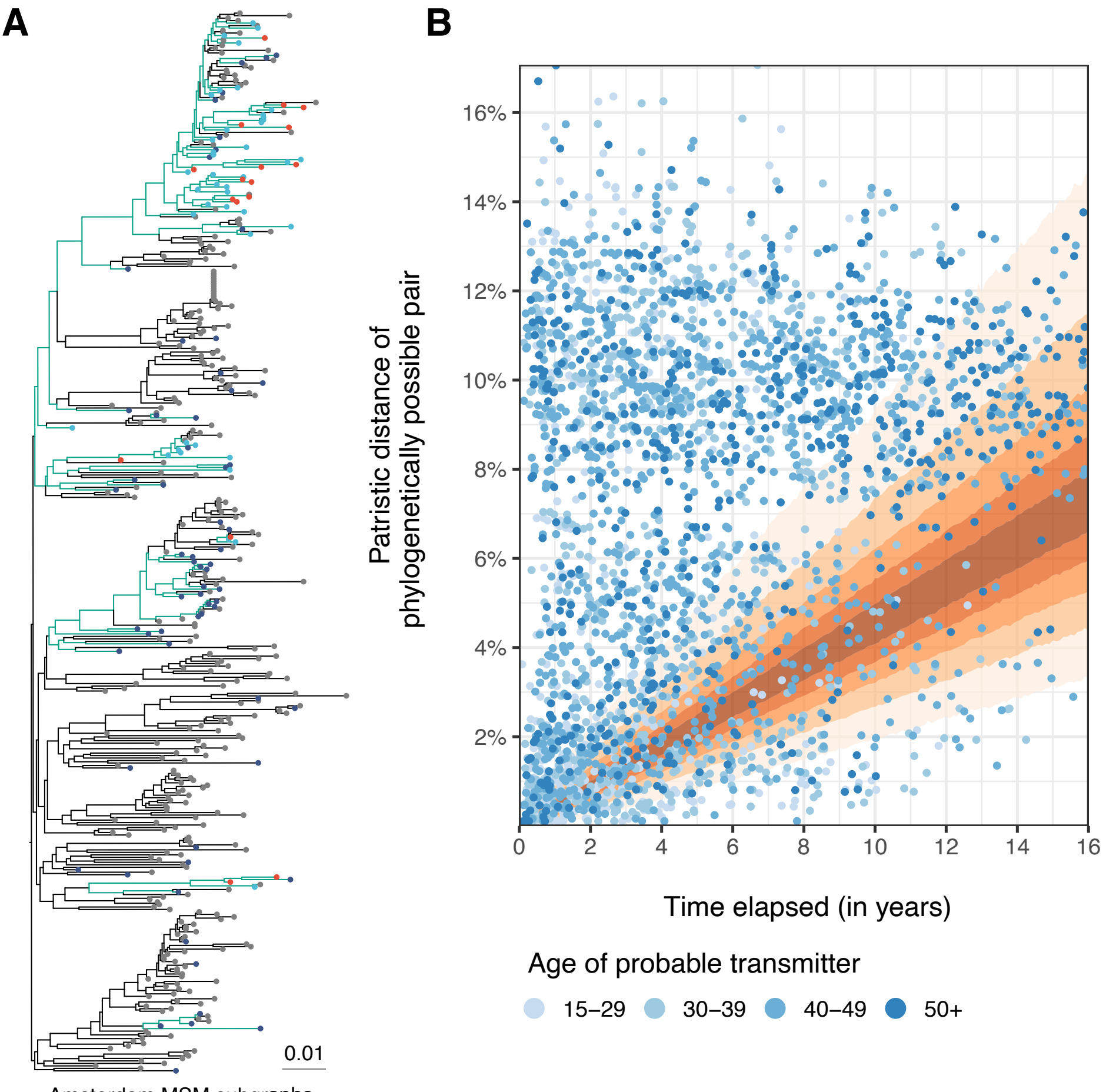

Figure 7: **Phylogenetic data from Amsterdam MSM.** A) Clade of subtype B among Amsterdam MSM. A clade is a subset of the phylogenetic tree including all descendents of an ancestral lineage. MSM subgraphs are identified by coloured branches. Members of MSM subgraphs with an infection date since 2010 have red tips; members who are a phylogenetically possible source for these recipients are have light blue tips; members who are neither a source nor a recipient have dark blue tips. B) Patristic distances and estimated time elapsed for phylogenetically possible transmission pairs of Amsterdam MSM.

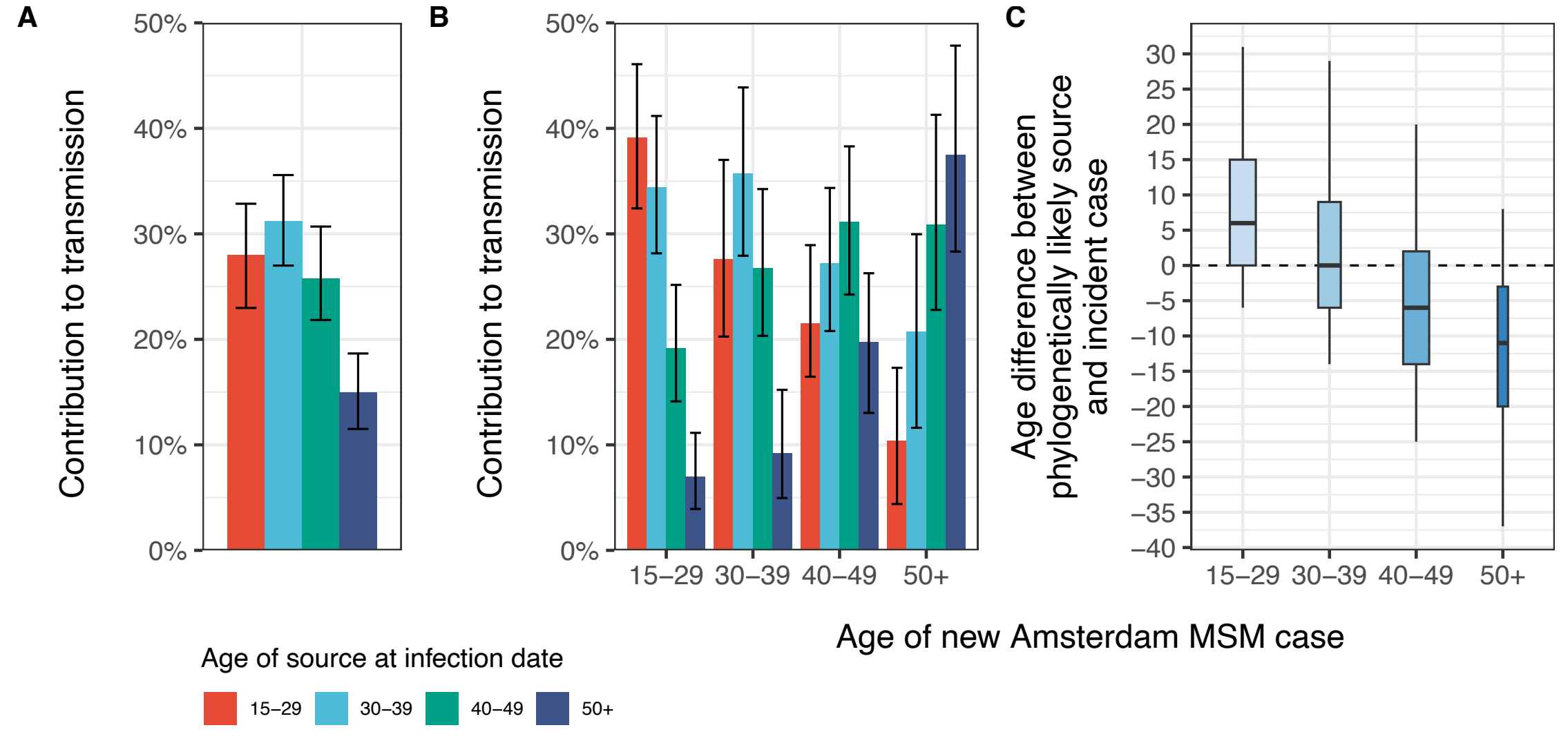

Figure 8: **Estimated age of Amsterdam MSM sources of transmission in Amsterdam MSM who acquired infection in 2010-2021.** A) Overall contribution of age groups to transmissions within Amsterdam MSM transmission networks in 2010-2021 (bar: posterior median, errorbar: 95% posterior credible intervals). B) Contribution of age groups to transmissions within each recipient age band of 15-29 years, 30-39 years, 40-49 years and 50 years and above. (bar: posterior median, errorbar: 95% posterior credible intervals). C) Estimated age difference between the sources and recipients in likely transmission pairs, with a positive age difference indicating older likely sources. The thick line within boxplots indicates the posterior median age difference, the box the posterior interquartile range, the whiskers the 2.5% and 97.5% posterior quantiles, and the widths of the boxplots is proportional to number of likely pairs who have a recipient in each recipient age band of 15-29 years, 30-39 years, 40-49 years and 50 years and above.

A central insight from our investigation is that the signal derived from combining molecular genetic distance measures with time elapsed remains ambiguous, and there is as in previous research[26,61,85] no way to identify the source of transmission with certainty. Simulations based on the PopART-IBM HIV transmission model demonstrate that the large majority of data points falling into the BMM signal component correspond to unlinked pairs of individuals, and thus are false signal that can introduce significant bias to population-level source attribution inferences. Additional features are necessary to reduce misclassification rates on the true, unknown transmission status between individuals in phylogenetically possible transmission pairs to a tolerable level for source attribution. These features may vary between pathogens and also between localised epidemics. For example, HIV suppression status will be most informative for estimating HIV transmission sources in populations with well-established treatment programs and regular monitoring of ART-response. For infectious diseases that can be cured, clinical data on cleared infections could similarly be used to ascertain when individuals were uninfectious. Diagnosis dates can be highly informative for pathogens whose serial intervals to the diagnosis date of the next infection are short[86–88]. Mortality data can be useful in narrowing down the potential sources of infection for life-long infectious diseases, and mobility data can be useful in excluding potential transmission sources where populations are highly mobile or suffer from displacement or armed conflict. From a methodological angle, these considerations imply that the vanilla BMM (13) is unlikely to provide accurate source attribution inferences, and we recommend using the covariate or random function BMMs (16)-(17), or mixtures of both. Other phylodynamic approaches based on inferring transmission trees from genomic data and infection times can be applied [89] using reversible jump MCMC which account for unsampled individuals. However, these are best suited to large, rapidly spreading outbreaks, whilst in the case of HIV there are generally many small concurrent transmission chains consisting of few individuals.

Here, we used the random function BMM to investigate the age groups that underpin continued spread of HIV within Amsterdam MSM transmission chains, excluding those for whom phylogenetic data indicates that they were likely infected by an individual not resident in Amsterdam. We found that no single age group drove transmission among MSM in Amsterdam between 2010-2021, though 30-39 year olds contributed the highest proportion of transmissions. Analysis of deep sequence data from six European countries participating in the BEEHIVE study, also identified MSM aged 30-39 to be the most likely source of infections across all age groups among phylogenetically linked pairs [90]. We found limited evidence of assortative mixing across all age groups, with the majority of infections in each of the recipient age bands 15-29 years, 30-39 years, 40-49 years and 50 years and above originating from outside these age groups respectively. Behavioural survey data collected from MSM in Amsterdam in 2008-2009 found disassortive mixing by age, particularly between casual partners [91], and phylogenetic evidence from MSM in Switzerland suggested an average overall age gap of 9 years between inferred transmission pairs [92]. We estimated age gaps of five years between MSM under 30 and their likely sources, and MSM over 40 were between five to twelve years younger than their likely sources. Age of sources have previously been found to depend on the age of the recipient at time of infection, with MSM under 30 in the European BEEHIVE study estimated to have a source on average 6 years older, and over 40 were estimated to have a source 8 years younger [90], cohesive with our findings. Other studies in the United States, focused on HIV transmission to young MSM, have also identified age gaps between recipients and older sources [93,94]. However a study among Tennessee MSM found evidence from phylogenetic analysis of more transmission between young MSM than from older partners [95]. Incorporating other covariates, White MSM have been estimated to have larger age gaps between young recipients and their sources than other races and ethnicities across the US [67], so age gaps may also be driven by population demographics. We did not quantify this in our study, however over 40% of new diagnoses in Amsterdam MSM are among individuals with a migration background [96], suggesting age gaps may be heterogeneous depending on ethnicity of incident cases.

Our statistical approach has a number of important limitations. First, underpinning the model are the infection dates of individuals, which are usually unknown but can be estimated. Existing methods for estimating infection time have in general been shown to suffer from individual-level uncertainty [38,75,81]. This is carried through to the time elapsed, which can lead to false transmission pair signal among truly unlinked pairs or lack of signal for actual transmission pairs, and thereby distort population-level flow estimates. However, accounting for infection time uncertainty in the application found no indications of substantial bias, though uncertainty intervals were larger. Second, we also do not adjust for incomplete sampling in the model, meaning the true source for an incident case may not be among the phylogenetically possible sources. In practical applications, for many pathogens and populations it is likely that a proportion of individuals do not have a sequence, due to being undiagnosed, or for reasons of study enrollment and consent [20]. For example, we were unable to fit the model to data from heterosexual phylogenies from Amsterdam, due to small sample sizes and low sequence coverage. Approaches have been developed for assessing whether datasets have sufficient samples to identify truly linked transmission pairs [97], and for accounting for incomplete sampling in source attribution, which goes beyond the scope of this study though could be incorporated in practice [98,99]. Thirdly, we focus on developing a parsimonious clock model to nest within the mixture model. There is limited data available for pairs with a time elapsed of less than one year and above 15 years to train the model; as a result, there is large uncertainty associated with inferences made for individuals with values outside this range. In addition, we consider a linear trend in the within-host evolutionary rate over time, though it is possible the rate decreases over a prolonged period[100]. Finally, we have demonstrated the application of this method to estimating sources of HIV. However, it could be applied to other fast evolving pathogens such as Hepatitis C and Ebola, if similar data from confirmed transmission pairs, or existing estimates of their evolutionary rate, are available[101,102]. Other predictors, with existing evidence of their association with known transmitters, may be used in place of age to inform the BMM mixture weights, and population-level transmission flows may be summarised by other covariates of interest[103].

In summary, this paper develops a mixture model framework for incorporating time since infection estimates and pathogen genomic data to estimate population-level sources of pathogen spread. We find that time since infection estimates are informative about characterizing transmission sources, both through reducing the number of potential sources for each new infection case, and by providing the data needed to interpret pathogen genomic data in the context of the signal derived from pathogen-specific molecular clocks. We also find that individual-level sources of transmission cannot be identified even with additional time since infection estimates, and demonstrate that false transmission pair signal is pervasive in realistic simulations of HIV spread as well as real-world data from Amsterdam. This prompted us to take a Bayesian approach that integrates out uncertainty in individual-level class labels, and estimates the relevance of additional covariates for source attribution through modelling generalized linear predictors of the mixture model mixing weights. The model has been principally developed to characterize HIV transmission flows, but is readily applicable to source attribution of other pathogens providing pathogen-specific molecular clocks can be specified.

## Acknowledgments

We thank the steering committee of the Amsterdam HIV transmission initiative and the Machine Learning & Global Health network for earlier comments on this work. The H-TEAM initiative is being supported by Aidsfonds (grant number: 2013169, P29701, P60803), Stichting Amsterdam Dinner Foundation, Bristol-Myers Squibb International Corp. (study number: AI424-541), Gilead Sciences Europe Ltd (grant number: PA-HIV-PREP-16-0024), Gilead Sciences (protocol numbers: CO-NL-276-4222, CO-US-276-1712, CO-NL-985-6195), and M.A.C AIDS Fund.

## Declaration of conflicting interests

AvS received funding for managing the ATHENA cohort, supported by a grant from the Dutch Ministry of Health, Welfare and Sport through the Centre for Infectious Disease Control of the National Institute for Public Health and the Environment; and received grants unrelated to this study from European Centre for Disease Prevention and Control paid to his institution. PR received grants unrelated to this study from Gilead Sciences, ViiV Healthcare and Merck, paid to his institution. GdB received honoraria to her institution for scientific advisory board participations for Gilead Sciences and speaker fees from Gilead Sciences (2019), Takeda (2018-2022) and ExeVir (2020-current). NP received grants unrelated to this study from ECDC and Gilead Sciences Hellas, paid to his institution; and received honoraria for presentations unrelated to this study from Gilead Sciences Hellas. OR received grants unrelated to this study from the NIH, and the Bill & Melinda Gates Foundation, paid to his institution.

## Funding

This study received funding as part of the H-TEAM initiative from Aidsfonds (project number P29701) and the Engineering and Physical Sciences Research Council (EP/X038440/1). The H-TEAM initiative is being supported by Aidsfonds (grant number: 2013169, P29701, P60803), Stichting Amsterdam Dinner Foundation, Bristol-Myers Squibb International Corp. (study number: AI424-541), Gilead Sciences Europe Ltd (grant number: PA-HIV-PREP-16-0024), Gilead Sciences (protocol numbers: CO-NL-276-4222, CO-US-276-1712, CO-NL-985-6195), and M.A.C AIDS Fund.

Supplementary Text to

# Bayesian mixture models for phylogenetic source attribution from consensus sequences and time since infection estimates: Supplementary Material

Blenkinsop et. al.

# S1 Supplementary Figures

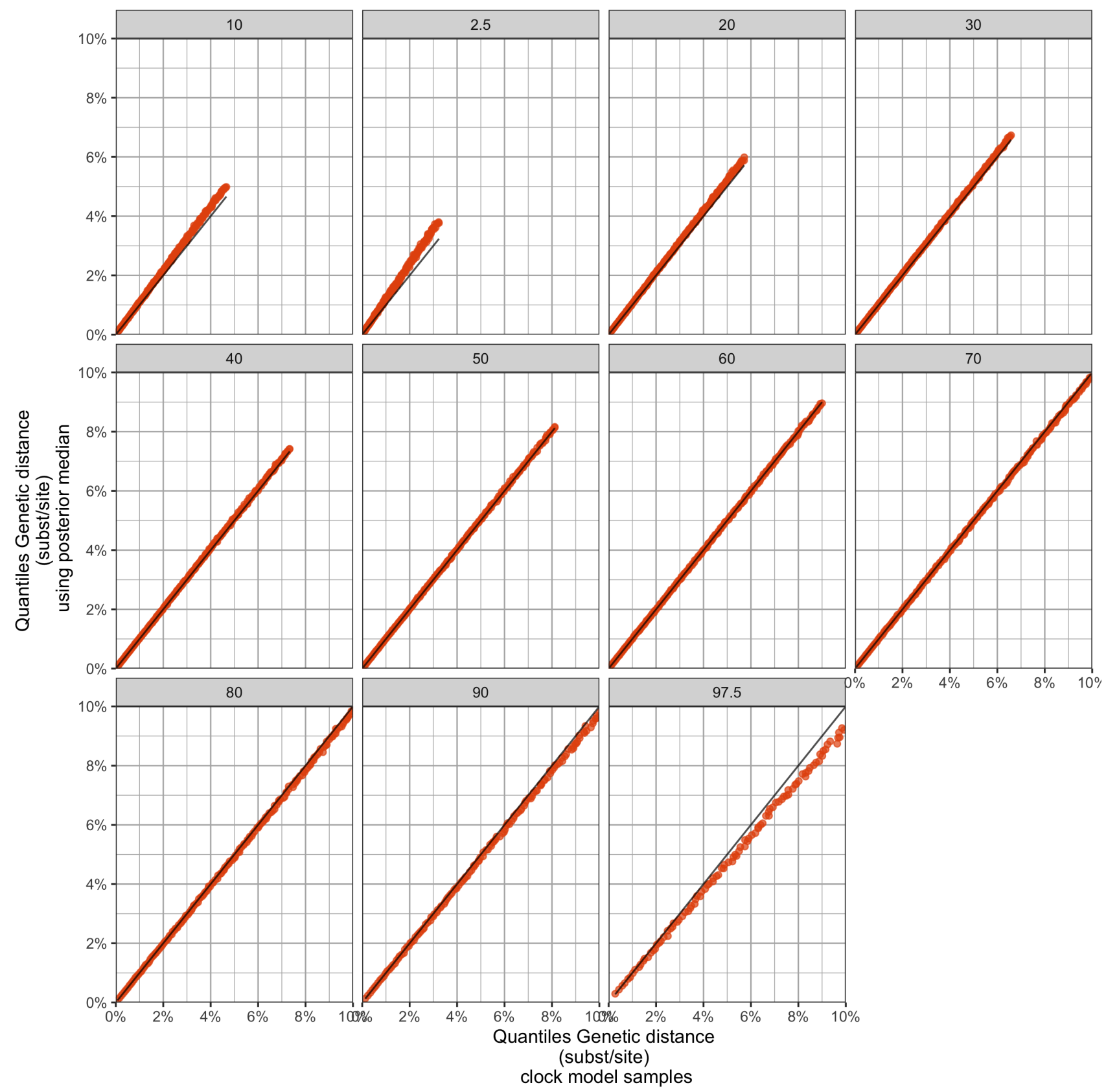

Figure S1: Quantile-Quantile plot for genetic distances predicted by the fitted molecular clock samples and using posterior median estimates to predict distances.

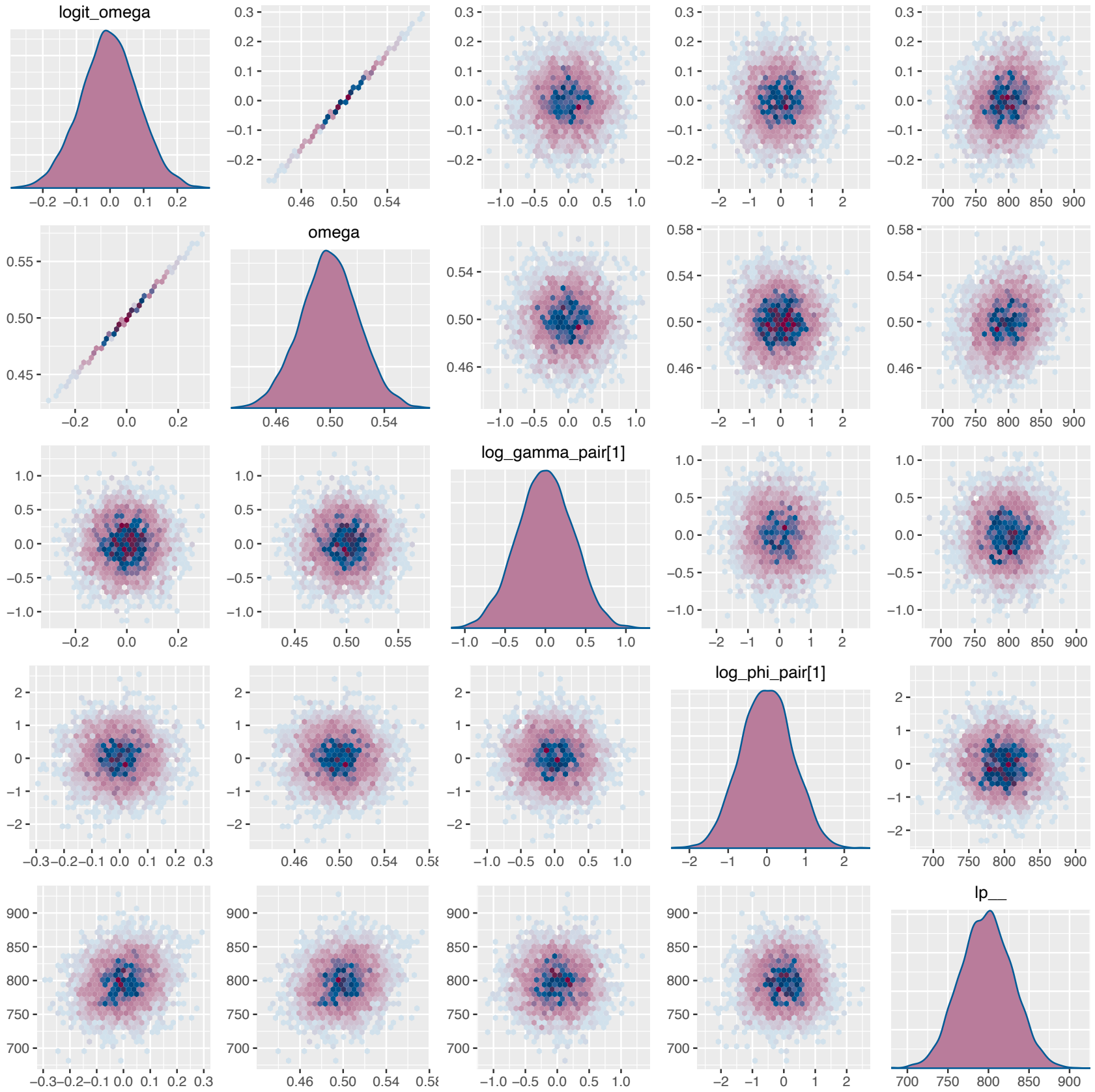

Figure S2: **Pairs plot of the joint posterior density of the vanilla BMM** (13) **parameters in the simulation study.** Omega is the mixing weight, logit_omega is the mixing weight (on the logit scale), log_gamma_pair[1] is the random effect of the evolutionary rate for one example pair in the data, log_phi_pair[1] is the random effect of the dispersion parameter for one example pair in the data.

# S2 HSGP random function hyper-parameters and tuning parameters

In the kernel for the HSGP function (17), hyper-parameters $\boldsymbol{\ell}$ and $\alpha$ are the characteristic length-scale and marginal variance of the kernel, respectively [1]. The length-scale determines the smoothness of the function, and the variance is a scaling factor, determining the deviation of values from their mean. We use the squared exponential covariance function, which has the property of being infinitely differentiable, so ensures ensures good smoothness

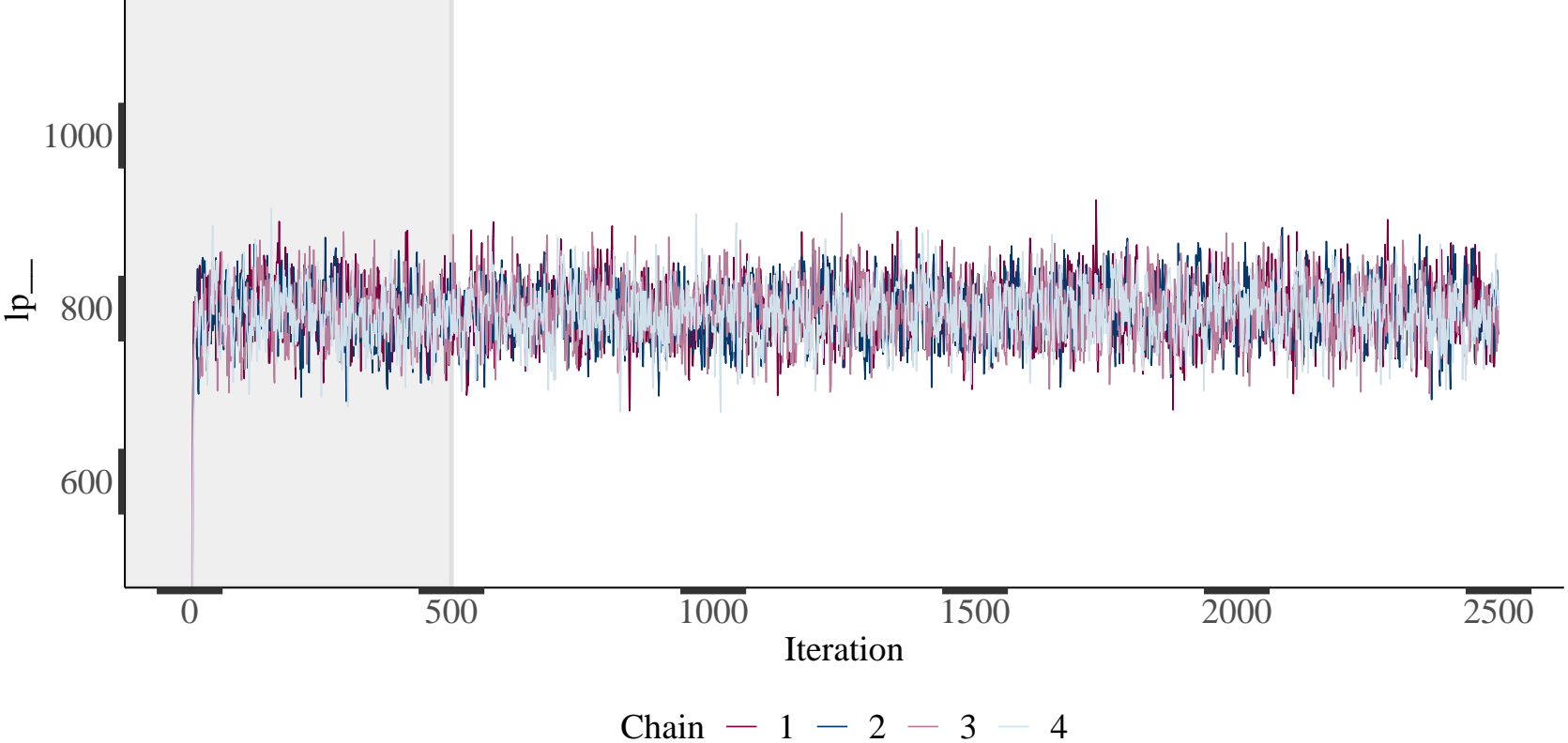

Figure S3: **Trace plot of parameter with the smallest effective sample size for the vanilla BMM in the simulation study.**

in the GP function [2].

The HSGP approximates the kernels with additional tuning parameters, $m$ and $B$. $m = m_1 \times m_2$ are the number of basis functions to approximate each of the kernels, which determines the accuracy of the GP approximation, and are chosen to balance accuracy and computational speed [3]. We chose $m_1 = m_2 = 24$. $B$ is a boundary factor which increases the shifted input domain of $\mathcal{A}$ (centred at zero) to $\Omega = [-L, L]$, defining the domain of the HSGP. Larger values of $B$ improve the the accuracy of the GP approximation, at the cost of computational speed; $B = 1.2$ was chosen using diagnostics to balance these criteria.

# S3 Evolutionary clock model

The model describing the patristic distances given time elapsed (8) assumes a linear relationship, since in the case of HIV there is substantial evidence to support linear intrahost evolutionary rates in early stages of infection [4, 5, 6, 7]. However the functional form may be flexible in other applications, in which there is evidence of non-linear relationships, and (8) can be adapted accordingly.

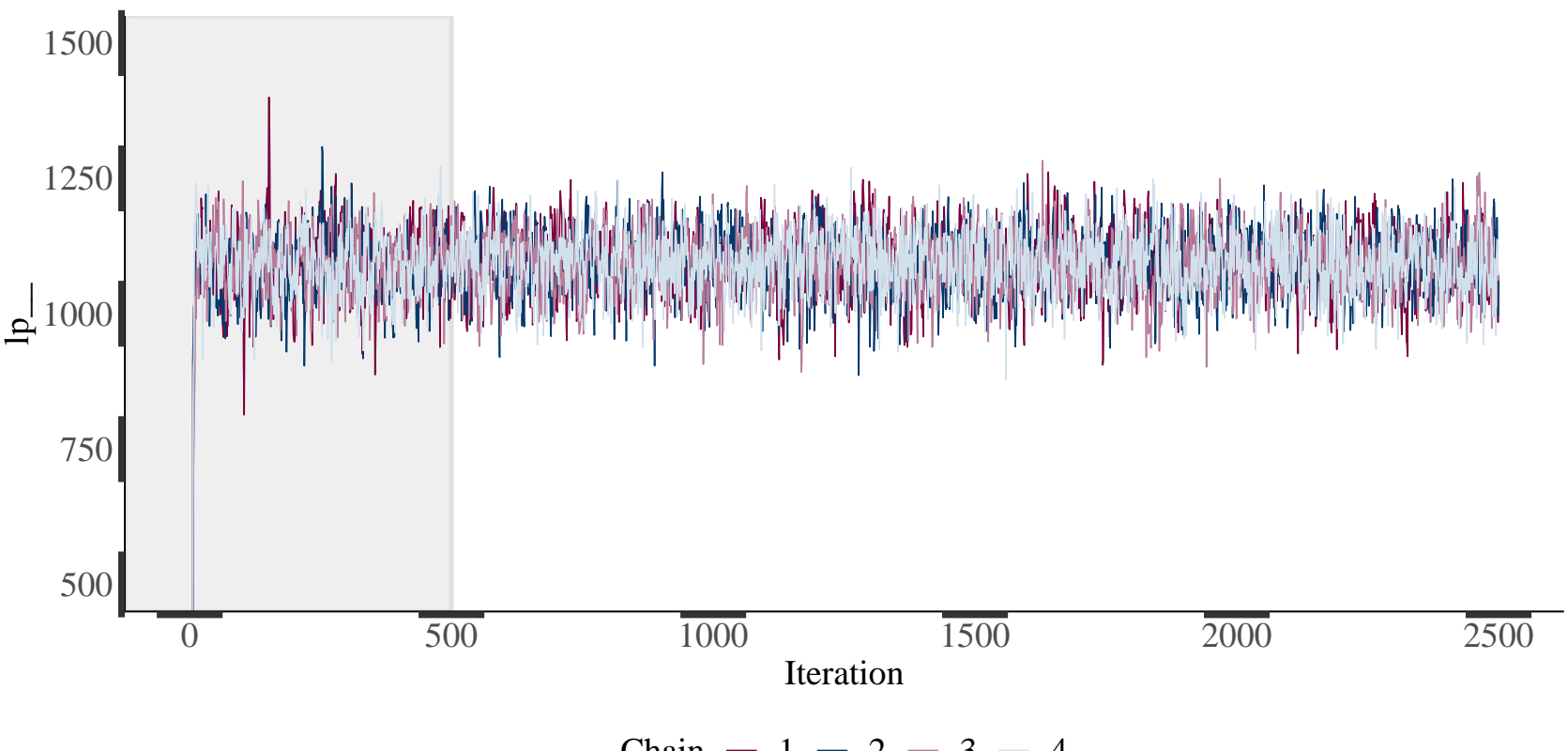

Figure S4: **Trace plot of parameter with the smallest effective sample size for HSGP BMM for Amsterdam MSM.**

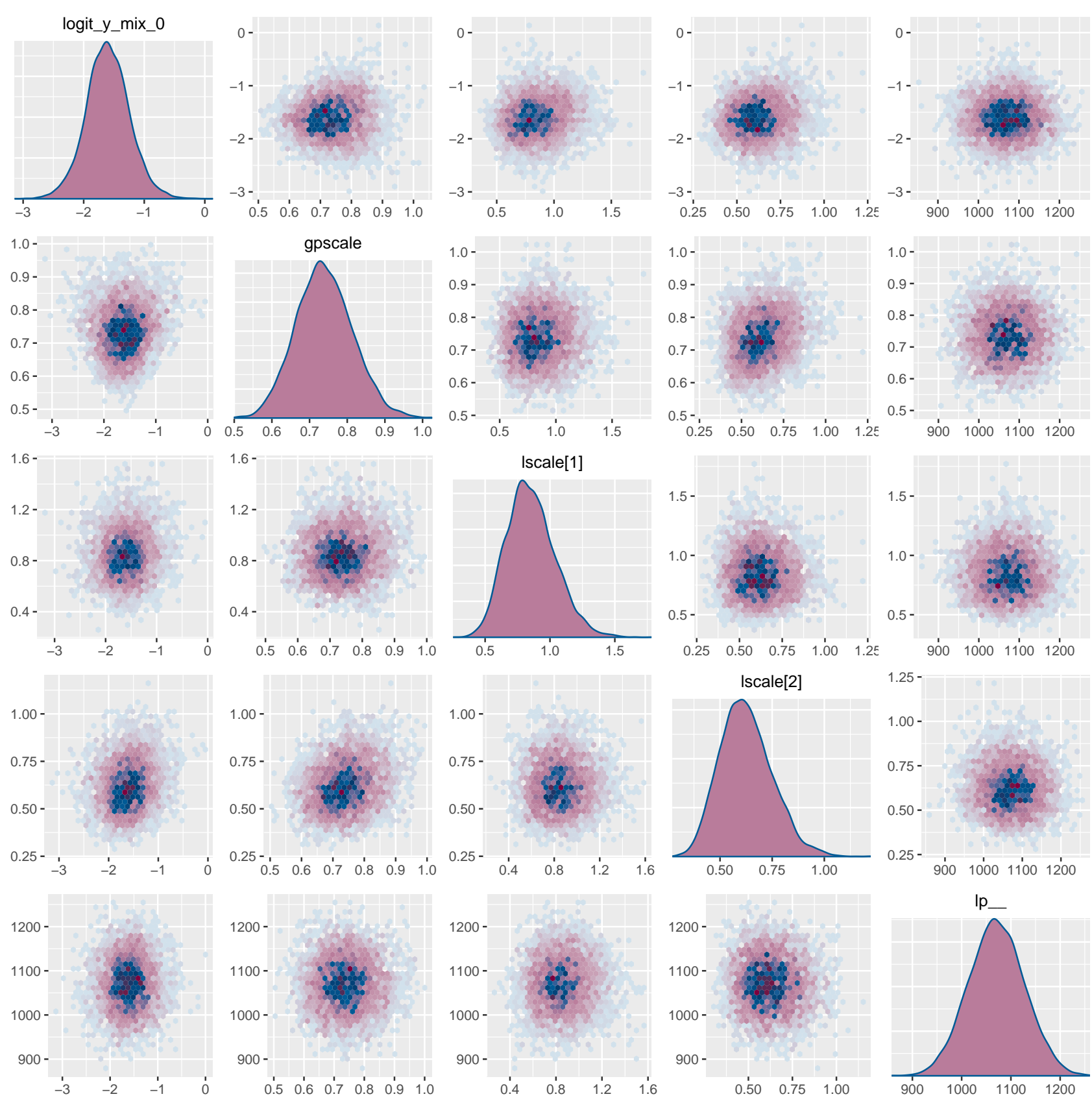

Figure S5: **Pairs plot of the joint posterior density of the HSGP BMM parameters for Amsterdam MSM.**

# S4 Epidemic simulation

## S4.1 Model

Transmission pairs were obtained from PopART-IBM [8], a discrete-time agent-based model developed contextually to the HPTN01 trial in Zambia and South Africa [9]. The model is able to simulate demographic processes, as well as HIV transmission dynamics and progression, and interventions at a community level. Interactions between people in the community and the surrounding area (not part of the trial) are also modelled. Only heterosexual partnerships are modelled, and the number of partners individuals may have in their lifetime depends not only on their age and sex, but also on the individual sex activity level, that is determined at birth. The model is informed by means of data collected prior and during the data, including demographic surveys, sexual surveys, data collected from community health care workers who delivered the intervention to households, and from health care facilities. Incidence and prevalence were measured on a cohort of 2000 people, representative of the population aged 18-44, followed up yearly.

PopART-IBM has a number of free parameters, that are calibrated to age-and-sex stratified data on incidence, prevalence, ART uptake, and viral suppression. Calibration is done in each community through Approximate Bayesian Computation algorithms. Transmission trees are obtained by running the model with the best-scoring set of free parameters from one random community, and considering transmission from 2006 onwards, in which at least the infected individual belonged to the community. To each infector-infectee pair we consider metadata including age, sex, set point viral load, cd4 counts of the infector. Simulations start in 1900, and HIV is introduced by infecting a random number of individuals each year between 1965 and 1970.

### S4.2 Parameters

The parameters that are calibrated refer broadly to three macro-categories: 1) initialization and HIV introduction; 2) sexual behaviour, including assortativity in partner choice and under-reporting; 3) HIV transmission and progression; 4) Cascade-care and ART effectiveness. The posterior distribution of such parameters is discussed in [8]. Table S1 report the chosen value of each parameter.

The simulation was initiated in 1900 with $N_0 = 4056$, with $32,217$ individuals alive by 1965 when the epidemic was seeded. 148 randomly selected individuals were infected between 1965-1970, and the simulation continued ran until December 2020. Overall, a total of $205,473$ individuals were simulated, with $34,961$ transmission events.

| Parameter | Value | Description |
|---|---|---|
| assortativity | 0.660 | sexual behaviour assortativity |
| c_multiplier | 3.22 | overall underreporting factor in number of partners |
| breakup_scale_multiplier | 1.56 | overall base partnership duration |
| average_annual_hazard | 0.092 | average annual hazard of transmission per individual |
| p_HIV_background_testing_pre2006 | 0.138 | background female rate of testing prior to 2006 |
| p_HIV_background_testing | 0.219 | background female rate of testing after 2006 |
| RR_HIV_background_testing_male | 0.924 | relative rate of testing for males |
| p_collect_cd4_test_result_nonpopart | 0.949 | background probability of entering the cascade care |
| log_seed_multiplier | 1.69 | annual number of seeds (1965-1970) |
| t_start_art_nonpopart | 0.444 | background time from infection to ART uptake |
| p_stays_virally_suppressed | 0.827 | overall probability of remaining virally suppressed after ART |
| p_stays_virally_suppressed_male | 0.989 | male probability of remaining virally suppressed after ART |
| RR_male_to_female_trans | 1.538 | relative rate of transmission from male to female |
| initial_low_risk_female | 0.456 | percentage of low sexual activity individuals - female |
| initial_low_risk_male | 0.487 | percentage of low sexual activity individuals - male |
| initial_med_risk_female | 0.826 | percentage of average sexual activity individuals - female |
| initial_med_risk_male | 0.758 | percentage of average sexual activity individuals - male |

Table S1: Free parameters for PopART-IBM.

### S4.3 Patient-level covariates

Some additional patient covariates were simulated to fit the BMM to the phylogenetically linked transmission pairs.

Time from infection to sequence sampling date, $\tau_i$ ($i = 1, \ldots, s$) were simulated from a Weibull distribution, with shape and scale parameters obtained by fitting a Weibull distribution to time-to-diagnosis estimates from Amsterdam MSM with the R package `fitdistrplus`:

$$\tau_i \sim \text{Weibull}(1.07, 2.89), \tag{S1}$$

as a proxy, assuming that individuals were sequenced shortly after diagnosis. Individuals sequence sampling date, $S_i$ $(i = 1, \ldots, s)$, was therefore defined by,

$$S_i = T_i + \tau_i. \tag{S2}$$

Time elapsed was calculated for all true transmission pairs, and non-transmission pairs using the infection dates and sequence sampling dates using (6). Next, genetic distances for true transmission pairs were simulated by first simulating random effects for each pair,

$$\log \gamma_{ij}^* \sim \mathcal{N}(0, \sigma_\gamma^m) \tag{S3}$$

$$\log \phi_{ij}^* \sim \mathcal{N}(0, \sigma_\phi^m), \tag{S4}$$

where $\sigma_\gamma^m$ and $\sigma_\phi^m$ denote the medians of the parameters from the molecular clock model. We next simulated the distances with the pair-specific parameters,

$$D_{ij} \sim Gamma(\alpha_{ij}^*, \beta_{ij}^*), \tag{S5}$$

$$\alpha_{ij}^* = \mu_{ij}\beta_{ij}, \tag{S6}$$

$$\mu_{ij} = (\gamma^m + \gamma_{ij}^*)T_{ij}^e, \tag{S7}$$

$$\beta_{ij}^{*-1} = \phi^m + \phi_{ij}^*, \tag{S8}$$

where $\gamma^m$ and $\phi^m$ are the posterior medians from the clock model. The distances for unlinked pairs were simulated uniformly,

$$D_{ij} \sim \text{Uniform}(0, 0.2). \tag{S9}$$

$$\tag{S10}$$

The ages of the sources, denoted by $x_{ij,1}$, and recipients, $x_{ij,2}$, on the infection date of recipient $j$, were simulated as follows,

$$x_{ij,1} \sim \text{LogNormal}(\log(30), \log(1.3)^2) \tag{S11}$$

$$x_{ij,2} \sim \text{LogNormal}(\log(x_{ij,1}), \log(1.25)^2), \tag{S12}$$

for $x_{ij,1}, x_{ij,2} \in [16, 75]$. The ages of the sources and recipients for unlinked pairs, were simulated uniformly,

$$x_{ij,1}, x_{ij,2} \sim \text{Uniform}(16, 75). \tag{S13}$$

To explore scenarios with fewer phylogenetically possible pairs per incident case, if $c$ are the average number of possible sources per recipient, $p = 1/c$ are the proportion of total pairs corresponding to true transmission pairs. After formulating all potential pairs and applying exclusion criteria, we randomly sampled non-transmission pairs to achieve $p = (50\%, 40\%, 30\%, 20\%, 10\%, 8\%)$, such that $c = (2, 2.5, 3.3, 5, 10, 12.5)$.

# S5 Application to Amsterdam

## S5.1 Generated quantities

To obtain the age differences between sources and recipients we define,

$$Z_{d,b}|\boldsymbol{X} = \sum_{i:a-b} \sum_{j \in b} \rho_{ij}|\boldsymbol{X}, \tag{S14}$$

where $a$ and $b$ are one-year age bands. We then aggregate over age groups of the recipient, $\tilde{b}$,

$$Z_{d\tilde{b}} = \sum_{b \in \tilde{b}} Z_{db}. \tag{S15}$$

We then define the flows from different one-year age gaps to age group $\tilde{b}$ by,

$$\delta_{d\tilde{b}} = Z_{d\tilde{b}} \Big/ \left( \sum_{d} Z_{d\tilde{b}} \right). \tag{S16}$$

We then obtain summary quantiles (25%, 50%, 75%) and minimum and maximum values for each recipient age group, $\tilde{b}$ for each monte carlo sample, then summarise by taking the median for each summary statistic across all samples.

# S6 Sensitivity analyses

## S6.1 Ages structure of actual and unlinked transmission pairs

To explore the impact of how well the age structure separates the actual transmission pairs from the unlinked pairs in the HSGP BMM, we simulated data in which the age structure of the unlinked pairs was more similar to the actual pairs. To disentangle the impact of how well the bivariate ages of sources and recipients informs their mixing probability from the estimated population-level sources, we considered a binary source category.

The ages of the sources, denoted by $x_{ij,1}$, and recipients, $x_{ij,2}$, on the infection date of recipient $j$, were simulated similarly as in (S11), with larger variance in the ages of the source to reduce the correlation between the ages within pairs,

$$x_{ij,1} \sim \text{LogNormal}(\log(30), \log(1.3)^2) \tag{S17}$$

$$x_{ij,2} \sim \text{LogNormal}(\log(x_{ij,1}), \log(1.8)^2), \tag{S18}$$

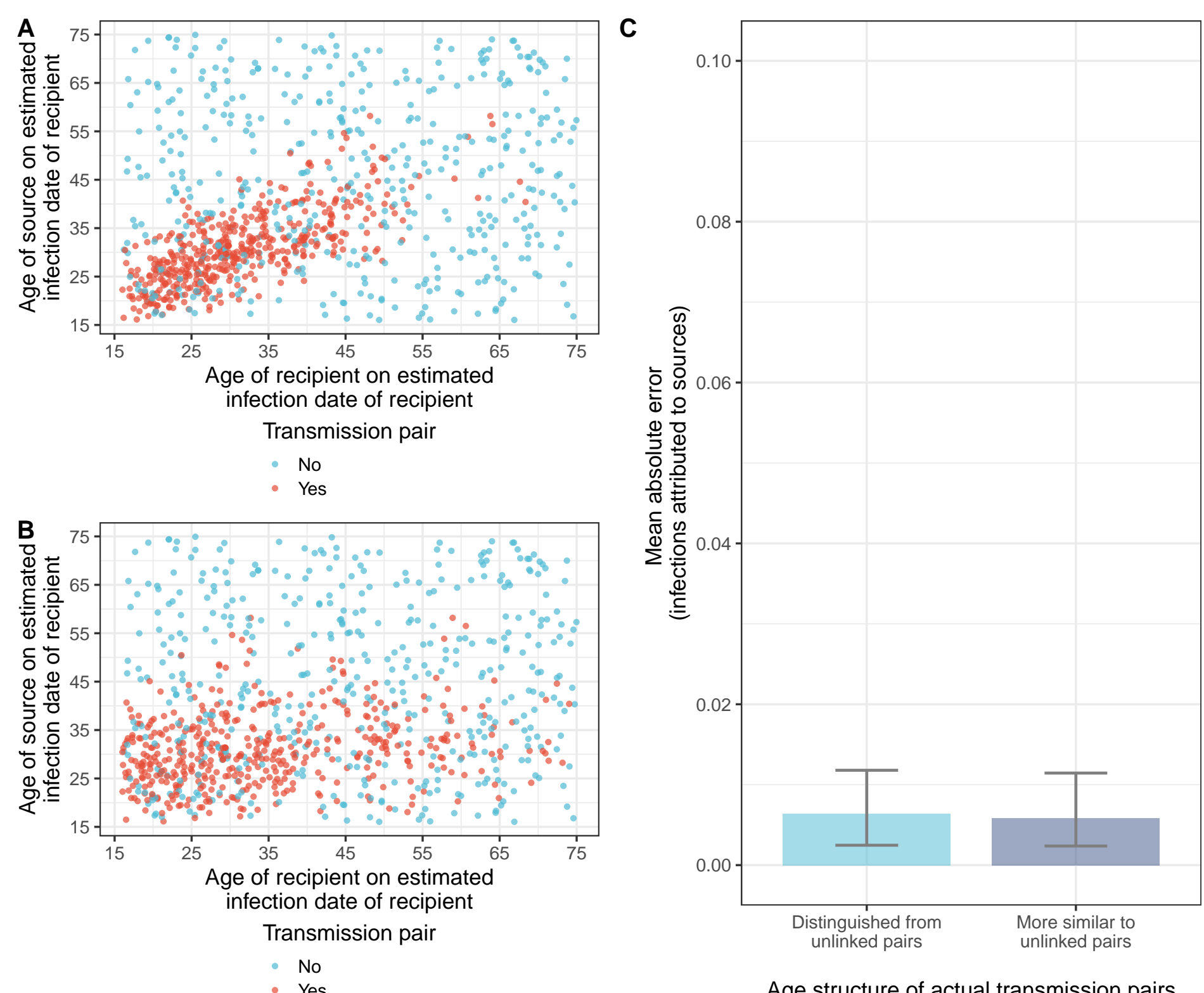

Figure S6: **MAE for binary source category for HSGP model under different age structures of sources and recipients**. (**A**) Structure of the simulated ages of sources and recipients in true transmission pairs and unlinked pairs under data generating procedure given by (S11). (**B**) Structure of the simulated ages of sources and recipients in true transmission pairs and unlinked pairs under data generating procedure given by (S17). (**C**) Posterior median estimates of mean absolute error (MAE) for both scenarios.

for $x_{ij,1}, x_{ij,2} \in [16, 75]$. The ages of the unlinked pairs were simulated uniformly, as before. Figure S6B shows the bivariate ages within linked and unlinked pairs, in comparison to Figure S6A from the primary results. The MAE for the estimated sources by five-year age groups was $< 1\%$ in both the model where the age structure of the actual transmission pairs was more distinguished from the unlinked pairs and more similar to the unlinked pairs.

| Linear predictor | Mean absolute error |
|---:|---:|
| Vanilla (no covariates) | 3.5% [3.2-3.8%] |
| Covariates on grouped ages of sources and recipients | 1.1% [0.7-1.5%] |
| 1D random function on age of source | 1.3% [0.8-1.8%] |
| 1D random function on age of recipient | 2.6% [2.3-2.9%] |
| 1D random functions on both age of source and recipient | 0.9% [0.6-1.3%] |
| 2D random function on age of source and recipient | 1.2% [0.8-1.6%] |

Table S2: Mean absolute error of models with different linear predictors on the mixture probability on the same number of pairs and average number of sources per incident case as the Amsterdam MSM data.

| Linear predictor | ELPD |
|---:|---:|
| Vanilla (no covariates) | 5144.0 |
| Covariates on grouped ages of sources and recipients | 5172.4 |
| 1D random function on age of source | 5178.6 |
| 1D random function on age of recipient | 5176.6 |
| 1D random functions on both age of source and recipient | 5164.9 |
| 2D random function on age of source and recipient | 5180.9 |

Table S3: Expected posterior log densities for BMM fitted to Amsterdam MSM with different linear predictors on the mixture probability.

### S6.2 Amsterdam results with different linear predictors

We ran simulations to quantify the mean absolute error for a similar configuration of pairs and average number of phylogenetically possible sources per incident case for various different models. Table S2 summarises their mean absolute errors.

We compared the final mixture model fitted to the Amsterdam MSM data to similar models with different linear predictors. Table S3 summarises the expected posterior log density (ELPD) for the four models.

### S6.3 Phylogenetically possible transmission pairs with patristic distances of zero

Seven Amsterdam MSM estimated to have seroconverted between 2010-2021 had a phylogenetically possible source with an identical sequence, leading to a patristic distance of zero. Since these appeared to be genuine distinct individuals, the patristic distance for these

| Age group of source | Estimated transmission sources from age group<br>Including pairs with zero distances* | <br>Excluding pairs with zero distances |
|---|---|---|
| 15-29 | 29.1% [26.5-32%] | 29.7% [27-32.8%] |
| 30-39 | 31.2% [28.6-34%] | 30.7% [27.9-33.4%] |
| 40-49 | 25.2% [22.5-27.9%] | 24.9% [22.1-27.6%] |
| 50+ | 14.4% [11.7-17.1%] | 14.8% [12-17.5%] |

* Setting their patristic distance to 0.077%

Table S4: Impact on estimated transmission sources among Amsterdam MSM in 2010-2021 from the HSGP BMM by excluding the seven pairs with a patristic distance of zero.

seven pairs was set to one mutation across the length of the alignment (0.077% substitution rate) before fitting the model. We carried out a sensitivity analysis excluding these pairs, which was found to have minimal impact on inferred sources of transmission by age group (Table S4).